\documentclass[11pt]{article}
\usepackage{graphicx} 
\usepackage{amssymb} 
\newcommand{\Comment}[1]{{}}
\usepackage{amsfonts,amsthm,slashed}
\usepackage[textwidth = 430 pt, textheight = 630 pt]{geometry}
\usepackage{color} 

\usepackage{setspace}
\usepackage{lipsum} 

\usepackage{cite}
\usepackage{url}

\usepackage{amsmath}

\newcommand\ignore[1]{}
\def\one{{\,\hbox{1\kern-.8mm l}}}

\def\Tr{{\rm Tr\, }}

\def\a{\alpha}\def\b{\beta}

\def\d{\partial}

\def\Tr{\mathop{\rm Tr}\nolimits}

\newcommand{\Cset}{{\,\,{{{^{_{\pmb{\mid}}}}\kern-.45em{\mathrm C}}}}}

\newcommand{\be}{\begin{equation}}
\newcommand{\bea}{\begin{eqnarray}}

\newcommand{\ee}{\end{equation}}
\newcommand{\eea}{\end{eqnarray}}

\providecommand{\lsim}{\lesssim}

\begin{document}


\renewcommand{\thefootnote}{\fnsymbol{footnote}}

\makeatletter
\@addtoreset{equation}{section}
\makeatother
\renewcommand{\theequation}{\thesection.\arabic{equation}}

\rightline{}
\rightline{}

\begin{center}
{\LARGE \bf{\sc Penrose limits and TsT for fibered $I$-branes}} 
\end{center} 
 \vspace{1truecm}
\thispagestyle{empty} \centerline{
{\large \bf {\sc Marcelo R. Barbosa ${}^{a,b},$}}\footnote{E-mail address: \Comment{\href{mailto:mr.barbosa@unesp.br}}
{\tt mr.barbosa@unesp.br}}
{\large \bf {\sc Horatiu Nastase${}^{a}$}}\footnote{E-mail address: \Comment{\href{mailto:horatiu.nastase@unesp.br}}
{\tt horatiu.nastase@unesp.br}}
{\bf{\sc and}}
{\large \bf {\sc Lucas S. Sousa${}^{a}$}}\footnote{E-mail address: \Comment{\href{mailto:santos.sousal@unesp.br}}{\tt santos.sousa@unesp.br}}
                                                        }

\vspace{.5cm}


\centerline{{\it ${}^a$Instituto de F\'{i}sica Te\'{o}rica, UNESP-Universidade Estadual Paulista}} 
\centerline{{\it R. Dr. Bento T. Ferraz 271, Bl. II, Sao Paulo 01140-070, SP, Brazil}}
\vspace{.3cm}
\centerline{{\it ${}^b$C.N. Yang Institute for Theoretical Physics,}}
\centerline{{\it S.U.N.Y. Stony Brook University, }} 
\centerline{{\it Stony Brook, NY 11794-3840, USA }}
 
\vspace{1truecm}

\thispagestyle{empty}

\centerline{\sc Abstract}

\vspace{.4truecm}

\begin{center}
\begin{minipage}[c]{380pt}
{\noindent 

In this paper we analyze a generalized "single-trace $T\bar T$" deformation, 
defined by a TsT transformation, of the fibered 
$I$-brane solution from \cite{Nunez2023}. We use the Penrose limit to try to
understand it, and we 
consider both the TsT followed by the Penrose limit, as well as the Penrose limit followed by TsT. 
We describe the dual spin chains obtained in field theory. In the first case we find that, indeed, the TsT 
transformation preserves solvability in a simple way, as in the standard $T\bar T$ case. In the second 
case, we have several options, but none is simple enough to be conclusive, however, one case gives us an 
asymptotically free and IR nontrivial field theory sector, and another a new parallelizable pp wave.

}
\end{minipage}
\end{center}

\vspace{.5cm}

\setcounter{page}{0}
\setcounter{tocdepth}{2}

    

\newpage

\tableofcontents
\renewcommand{\thefootnote}{\arabic{footnote}}
\setcounter{footnote}{0}

\linespread{1.1}
\parskip 4pt

\section{Introduction}

In the Penrose limit, the AdS/CFT correspondence \cite{Maldacena_1999} becomes simpler and 
easier to understand, besides giving us information about strings, not just supergravity 
\cite{Berenstein_2002}. That is why it has become a very good tool for analyzing holographic pairs 
that are not so well understood, see for instance 
\cite{Araujo_2017,Itsios:2017nou,Nastase2022,Barbosa2024a,Barbosa2024}.


An intriguing system of holographic pair is the "$I$-brane" \cite{Green1996}, which is also a particularly 
useful system to discuss anomalies for brane configurations. They are supergravity configurations 
consisting of intersections of two $p$-branes (or a $p-$brane and a $q-$brane, with $q \neq p$) over $n$ 
dimensions, with these spatial dimensions being interpreted as a $n$-brane. Specifically, \cite{Itzhaki2006} 
discussed the configuration of a 2-dimensional 
intersection of 5-branes in type IIB theory, that from the point of view of closed strings gains an 
extra dimension. In \cite{Nunez2023}, 2 types of holographic duals for the $2,5$-brane configuration 
were found, one singular, and one 
non-singular, the latter for a twisted compactification of the $I$-brane theory, i.e., a "fibered $I$-brane" 
system. This was also analyzed with the Penrose limit, finding the dual spin chain, in \cite{Barbosa2024}. 

But as a 2-dimensional theory (at least in terms of the open string construction), one natural question 
to ask is: can one characterize its $T\bar T$ deformation, which is known to preserve integrability
and to be in general solvable? The $T\bar T$ deformation 
\cite{Zamolodchikov:2004ce,Smirnov:2016lqw}, or more precisely its "single-trace" version
\cite{giveon2020tbartlst,Giveon:2017myj}, that has a normal holographic dual,
was argued to be described by a TsT transformation 
\cite{Lunin_2005,Frolov_2005} in \cite{Araujo_2019}.
Moreover, as we saw in \cite{Nastase2022,Barbosa2024}, acting with the $T\bar T$ deformation 
before or after the Penrose limit implies different results, so we must do this as well. 

In this paper then, we consider the $T\bar T$ (described by a TsT in the holographic dual) deformation, 
followed by a Penrose limit, and describe the dual spin chain, and we also consider the TsT
deformation(s) of the Penrose limit, and general structure of the corresponding spin chain(s), but 
in this latter case the 
analysis is less conclusive. 

The paper is organized as follows. In section 2 we review the $I$-brane background(s). In section 3 
we consider the TsT transformation of it, followed by the Penrose limit, and the analysis of the dual 
field theory and spin chain limit. In section 4 we consider the TsT transformation of the (previously 
obtained) Penrose limit of the $I$-brane background, first in the directions relevant to the 
case in section 2, then in other directions. In section 5 we try to interpret the results in terms of 
$T\bar T$ deformations, and in section 6 we conclude. The Appendix contains a review of the 
supergravity solutions, the TsT transformation rules, and the Page charges.

\section{Review of $I$-brane background(s)}

In this section, we will provide a brief review of the nonsingular fibered $I$-brane \cite{Itzhaki2006}
background, first defined and studied in  \cite{Nunez2023}.

The background describes a $(1+1) $-dimensional theory obtained when two stacks of $D5$-branes intersect 
along two spacetime directions, the "$I$-brane" configuration. Effectively, the theory acquires an extra 
worldvolume direction, as explained in \cite{Itzhaki2006}, thus 
being described by an $(2+1)$-dimensional theory.

In \cite{Nunez2023}, an IR completion for this theory was introduced by a 
twisted compactification on a shrinking cycle, obtaining a "fibered $I$-brane".

The supergravity background contains the metric $G_{\mu\nu}$, a dilaton $\Phi$, and an $RR$ field $C_2$. 
In the string frame they can expressed as
\bea
ds_{st}^{2}&= & r\left\{ -dt^{2}+dx^{2}+f_{s}(r)d\varphi^{2}+\frac{4}{r^{2}f_{s}(r)}dr^{2}
+\frac{2}{e_{A}^{2}}\left[\hat{\omega}_{1}^{2}+\hat{\omega}_{2}^{2}+\left(\hat{\omega}_{3}
-e_{A}Q_{A}\zeta(r)d\varphi\right)^{2}\right]\right.\cr
 && \left.+\frac{2}{e_{B}^{2}}\left[\tilde{\omega}_{1}^{2}+\tilde{\omega}_{2}^{2}+\left(\tilde{\omega}_{3}
 -e_{B}Q_{B}\zeta(r)d\varphi\right)^{2}\right]\right\} ;\cr
C_{2}&= & \psi_{A}\left(\frac{2Q_{A}}{e_{A}}\zeta^{\prime}(r)dr\wedge d\varphi-\frac{2}{e_{A}^{2}}\sin\theta_{A}
d\theta_{A}\wedge d\phi_{A}\right)+\frac{2}{e_{A}}\cos\theta_{A}Q_{A}\zeta(r)d\varphi\wedge d\phi_{A}\cr
 && +\psi_{B}\left(\frac{2Q_{B}}{e_{B}}\zeta^{\prime}(r)dr\wedge d\varphi-\frac{2}{e_{B}^{2}}\sin\theta_{B}
 d\theta_{B}\wedge d\phi_{B}\right)+\frac{2}{e_{B}}\cos\theta_{B}Q_{B}\zeta(r)d\varphi\wedge d\phi_{B};\cr
\Phi&= & \log r.
\label{eq:metric original}
\eea
The functions $f_s(r)$ and $\zeta(r)$ are given by
\begin{equation}
\begin{aligned}
f_s(r) & =\frac{e_A^2+e_B^2}{2}-\frac{m}{r^2}-\frac{2\left(Q_A^2+Q_B^2\right)}{r^4} \equiv \frac{e_A^2+e_B^2}{2 r^4}\left(r^2-r_{+}^2\right)\left(r^2-r_{-}^2\right), \\
\zeta(r) & =\frac{1}{r^2}-\frac{1}{r_{+}^2}, \quad r_{ \pm}^2=\frac{m \pm \sqrt{m^2+4\left(Q_A^2+Q_B^2\right)\left(e_A^2+e_B^2\right)}}{e_A^2+e_B^2}.
\end{aligned}
\label{function f zeta}
\end{equation}
Here $\hat \omega_i, \tilde \omega_i$ are the 
Maurer-Cartan forms for $\mathrm{su}(2)$, given by
    \begin{eqnarray}
& & \hat{\omega}_1= \cos {\psi_A} d{\theta_A} +\sin{\psi_A}\sin\theta_A d{\phi_A},\;\;\;\;\;
 \tilde{\omega}_1= \cos {\psi_B} d {\theta_B} +\sin {\psi_B}\sin\theta_B d{\phi_B},\nonumber\\
& & \hat{\omega}_2= -\sin{\psi_A} d{\theta_A} +\cos {\psi_A}\sin\theta_A d{\phi_A},\;\;
\tilde{\omega}_2= -\sin {\psi_B} d{\theta_B} +\cos {\psi_B} \sin\theta_B d {\phi_B},\nonumber\\
& & \hat{\omega}_3= d{\psi_A} +\cos{\theta_A}d{\phi_A},\;\;\;\;\;\;\;\; ~~~~~~~~~\;\;\;\;\; 
\tilde{\omega}_3= d{\psi_B} +\cos {\theta_B}d {\phi}_B.\nonumber
    \end{eqnarray}

The supersymmetry of this background was studied in \cite{Nunez2023}, where it was 
found that it preserves four 
supersymmetries when the parameters $e_{A,B},m,Q_{A,B}$ satisfy $m=0,\ e_AQ_B=\pm e_BQ_A$.
The background is confining, and the space ends smoothly (without a singularity) in the IR, at $r=r_+$, 
the largest of the solutions of $f_s(r)=0$. This is possible because of the compactification on the $S^1$ 
defined by $\varphi$. Since $e^\Phi$ will become large at some $r_*$, the above D5-D5 solution 
wrapped on $S^3$ and $\tilde S^3$ (as defined by the large $r$ region), 
needs to be S-dualized to an NS5-NS5 one, see \cite{Nunez2023,Itzhaki2006}.

The Penrose limits and the corresponding spin-chains of the above background were studied 
in \cite{Barbosa2024}. 
The Penrose limit in different directions was shown to be the same and has the form of a 
parallelizable plane wave,
\bea
d s^2 & =&2 d u d v-(d u)^2\left[a_1^2\left(x_1^2+x_2^2\right)+a_2^2\left(x_3^2+x_4^2\right)
+a_3^2\left(x_5^2+x_6^2\right)+a_4^2\left(x_7^2+x_8^2\right)\right]\cr
&&+\sum_{i=1}^8 d x_i^2 \cr
H & =&d u \wedge\left(2 a_1 d x_1 \wedge d x_2+2 a_2 d x_3 \wedge d x_4+2 a_3 d x_5 
\wedge d x_6+2 a_4 d x_7 \wedge d x_8\right) \Rightarrow \cr
B & =&d u \wedge\left[a_1\left(x_1 d x_2-x_2 d x_1\right)+a_2\left(x_3 d x_4-x_4 d x_3\right)
+a_3\left(x_5 d x_6-x_6 d x_5\right)\right.\cr
&&\left.+a_4\left(x_7 d x_8-x_8 d x_7\right)\right].
\label{pp wave original}
\eea

In the supersymmetric case commented on before, we have $a_1=a_4=0$ and $a_2=a_3$. 
In this case, we have an enhancement of supersymmetries, and the total SUSY charges 
preserved are 24, as can be found in \cite{Sadri:2003bk}, where the general case of parallelizable plane
waves was studied.


\section{TsT transformation ($T\bar T$ deformation) plus Penrose limit}

As we described previously, the "single-trace" $T\bar T$ deformation of the 2-dimensional 
field theory corresponds in the holographic dual to a TsT transformation. 

Therefore in this section we start by studying 
the TsT (T-duality, shift, T-duality) transformation of the twisted solution
 (\ref{eq:metric original}), called background II in  \cite{Nunez2023} and then its Penrose limit.\footnote{Note
 that in \cite{Asrat:2023yzy} the combination of TsT and Penrose limit was also considered in the case 
 of NS-NS $AdS_3\times S^3\times T^4$, giving the original single-trace $T\bar T$ deformation.}


\subsection{TsT transformation ($T\bar T$ deformation) in gravity and gauge theory}

We start by doing the TsT transformation, by choosing a pair of isometries in 
 (\ref{eq:metric original}). We will do the T-duality in the $x$-direction and the shift in the 
time direction, $t$.

The general rules for T duality (both for the NSNS and the RR fields) are known from the literature, 
see \cite{BUSCHER198759}, \cite{Itsios_2012}, \cite{Meessen_1999}, but for the reader's convenience, 
the general rules are given in the Appendix \ref{appendicea2}. 
However, as in \cite{Hashimoto:1999ut,Maldacena:1999mh,Nastase2023}, one needs 
to first Wick rotate to Euclidean space, then to the TsT transformation, and then Wick rotate back, 
in order not to get an apparent singularity in the  metric.
Since the TsT transformation acts on $x$ and $t$, in the intermediate steps we rewrite only the parts 
of the metric affected by these transformations. 
Performing first a T-duality along $x$, we obtain
\bea
    ds^2&=&+rdt^2+\frac{1}{r}dx^2+\cdots.\cr
    \Phi&=&\frac{1}2{\ln{r}}\;,
\eea
with $B=0$. Next, we do a shift $t\rightarrow t+\eta x$,
\begin{equation}
    ds^2=+rdt^{2}+
    2\eta rdtdx+\left(\frac{1+\left(r\eta\right)^{2}}{r}\right)dx^{2}+\cdots\;,
\end{equation}
with all other fields remaining the same at this step. 

Finally we T-dualize back in $x$, and Wick rotate back to Minkowski space. 
The NSNS part of the background is
\begin{equation}
\begin{aligned}ds_{st}^{2}= & \left(\frac{r}{1+\left(r\eta\right)^{2}}\right)
\left(-dt^{2}+dx^{2}\right)+r\left\{ f_{s}(r)d\varphi^{2}+\frac{2}{e_{A}^{2}}\left[\hat{\omega}_{1}^{2}
+\hat{\omega}_{2}^{2}+\left(\hat{\omega}_{3}-e_{A}Q_{A}\zeta(r)d\varphi\right)^{2}\right]\right.\\
 & \left.+\frac{4}{r^{2}f_{s}(r)}dr^{2}+\frac{2}{e_{B}^{2}}\left[\tilde{\omega}_{1}^{2}+\tilde{\omega}_{2}^{2}+
 \left(\tilde{\omega}_{3}-e_{B}Q_{B}\zeta(r)d\varphi\right)^{2}\right]\right\} ,\\
B= & +\left(\frac{r}{1+\left(r\eta\right)^{2}}\eta r\right)dx\wedge dt\\
\Phi= & \frac{1}{2} \ln\left(\frac{r^{2}}{1+\left(\eta r\right)^{2}}\right).
\end{aligned}
\label{eq: final back}
\end{equation}

Since the TsT transformation is a {\em symmetry} of the string theory, and it acts on the 
field theory coordinates, the modification should not change the solvability of the theory, a fact consistent
with $T\bar T$. Also, since 
the deformation is ${\cal O}(r^2)={\cal O}(r^{\Delta-d})$ (with $r$ the standard radial coordinate
and $d=2$), 
this gives a $\Delta=4$ deformation, just like $T\bar T$ is supposed to be in $d=2$. Also, since it is 
the metric that is deformed, that couples to $T_{\mu\nu}$, it should be something like $\sim \det T_{\mu\nu}
\sim T\bar T$. We will say more about this later, but this is similar to what was obtained in 
\cite{Nastase2023}, just that now in the better understood $d=2$ case, for instance 
since \cite{Araujo_2019} rigorously proved that TsT equals $T\bar T$ for the NS-NS $AdS_3\times S^3
\times T^4$ case. Also, exactly like in that case, we note that, since $\Delta>d$, the deformation corresponds
to an {\em irrelevant} operator, thus modifies the UV of the theory, which is not a CFT anymore (as it 
happens for any 
$T\bar T$ deformation).

The RR fields are also transformed under TsT, and we get a new $RR$ field $C_4$, with the same $C_2$,
\begin{equation}
    \begin{aligned}
        C_{2}= & \psi_{A}\left(\frac{2Q_{A}}{e_{A}}\zeta^{\prime}(r)dr\wedge d\varphi-\frac{2}{e_{A}^{2}}\sin\theta_{A}d\theta_{A}\wedge d\phi_{A}\right)+\frac{2}{e_{A}}\cos\theta_{A}Q_{A}\zeta(r)d\varphi\wedge d\phi_{A}\\
 & +\psi_{B}\left(\frac{2Q_{B}}{e_{B}}\zeta^{\prime}(r)dr\wedge d\varphi-\frac{2}{e_{B}^{2}}\sin\theta_{B}d\theta_{B}\wedge d\phi_{B}\right)+\frac{2}{e_{B}}\cos\theta_{B}Q_{B}\zeta(r)d\varphi\wedge d\phi_{B};\\
        C_{4}=&B\wedge C_2.
    \end{aligned}
\end{equation}

It is also important to impose that $dC_4-H\wedge C_{2}=F_{5}$ is self-dual. 
This  is imposed (since type IIB supergravity doesn't have an "usual action" for $C_4$, \cite{Pasti_1997})
by  acting with the projector $P_{+}=(1+\star_{10})$ on $F_5$,
\begin{equation}
    \tilde{F_{5}}=(1+\star_{10})B\wedge F_{3}.
\end{equation}

The background generated is, as expected \cite{Lunin_2005}, \cite{Frolov_2005}, a supergravity solution of 
type IIB, as can be checked using the equations in the Appendix \ref{appendicea1}. The field content in this 
supergravity setup implies the presence of D3-branes that couples to the $C_4$ field, D5-branes coupling magnetically to $C_2$ and F1-strings, coupling electrically with the 
Kalb-Ramond B field, with these coupling generating fluxes that are quantized, \cite{Page:1983mke}, 
\cite{marolf2000chernsimonstermsnotionscharge}. For the D5-branes, they are the same as before deforming, 
$N_{A}=N_{A, D5}$ and $N_{B}=N_{B, D5}$. To obtain the Page charges,\footnote{To remind the 
reader, the Page charge of D$p$-branes, 
$Q_{\rm Page, p}=\frac{1}{(2\pi l_s)^{7-p}}\int_{\Sigma_{8-p}}\hat F_{8-p}$, where $\hat F_p=[e^{-B\wedge}
\tilde F]_p$, is the only one that is quantized, but it is not gauge invariant (changes under large gauge 
transformations), while the Maxwell charge, $Q_{\rm Max,p}=\frac{1}{(2\pi l_s)^{7-p}}\int\tilde F_{8-p}$, 
is conserved and gauge invariant, but not quantized, where $\tilde F_p$ has a nontrivial Bianchi 
identity involving $H_3=dB$, for instance $d\tilde F_4=H_3\wedge \tilde F_2$, and the brane charge is one 
that is gauge invariant, but not conserved nor quantized, and is the integral of the source (breaking) of the 
Bianchi identity, for instance $\int_{M^5} (d\tilde F_4-H_3\wedge \tilde F_2)$. For the F1, this is a bit 
less well defined, but we will follow \cite{ferko2025holographynullboundaries} for the analog of the 
(quantized) Page charge. }
we follow the line of reasoning 
of \cite{ferko2025holographynullboundaries}, and with the calculations done in detail in 
Appendix \ref{appendicea3}, we get
\begin{equation}
    Q_{F_{1}}\sim\int \star_{10}\left ( e^{-2 \phi} H \right ) - C_2 \wedge \star (B \wedge F_3) ;\ \ Q_{D_{3}} 
    \sim \int \star (B \wedge F_3)\;,
\end{equation}

We start by computing the D3 brane density,
\bea
\label{b star f}
    \star (B \wedge F_3) &=& 
  4 \eta \Bigg[ 
    \,  \frac{r e_B \sin \theta_A}{e_A^3} \, d\psi_A \wedge dr \wedge d\phi \wedge d\theta_A \wedge d\phi_A \cr
    &&+
    \frac{ r e_A \sin \theta_B}{e_B^3 } \, d\phi \wedge dr \wedge d\psi_B \wedge d\theta_B \wedge d\phi_B \cr
   & &+2  Q_B \cdot \left( \frac{\sin \theta_A \sin \theta_B}{ e_A^3 e_B^2 } \right) 
  d\psi_A \wedge d\theta_A \wedge d\phi_A \wedge d\theta_B \wedge d\phi_B \cr
    &&- 2  Q_A \cdot \left( \frac{\sin \theta_A \sin \theta_B}{ e_B^3 e_A^2 } \right)
     d\psi_B \wedge d\theta_A \wedge d\phi_A \wedge d\theta_B \wedge d\phi_B
    \Bigg].
\eea

Choosing two cycles orthogonal to the radial direction and the time, 
one localized at $\psi_A$ and the other at $\psi_B$. From this, we obtain $N_3$,
\bea
\label{b star f}
    N_3 &=& \frac{8 \eta}{e_A e_B} \int (e_A Q_A \sin \theta_A \sin \theta_B  d \theta_A \wedge d \theta_B 
    \wedge d \phi_A \wedge d \phi_B \wedge d \psi_B\cr
    && - e_B Q_B  \sin \theta_A \sin \theta_B d \theta_A \wedge d \theta_B \wedge d \phi_A \wedge 
    d \phi_B \wedge d \psi_A ) \cr
    &=&  \frac{2 \text{Vol}_{S_3}^2 \eta}{\pi e_A^3 e_B^3} (e_A Q_A - e_B Q_B)\;,
\eea
where $\text{Vol}_{S_3} = \int \sin \theta d \theta d \phi d \psi$, in terms of the Euler angles. 
Notice that the exact volume of $S_3$ would be normalized by $\frac{1}{8}$.

For the fundamental string we must also choose two-cycles. We choose them such that  we have fundamental strings localized at 
$\psi_A = e_B^2$ and at $\psi_B = e_A^2$ (up to normalization), and extended over the $x$ direction. 
The reason for this choice is because we obtain a finite $N_1$. 
The total density of the $F_1$ charge is, for large $r$,
\begin{equation}
\begin{aligned}
    \lim_{r \rightarrow \infty} \rho_1 = \frac{-8 \eta r \sin \theta_A \sin \theta_B}{e_A^3 e_B^3} ( e_B^2 \psi_B d r \wedge d \theta_A \wedge d \theta_B \wedge d \varphi \wedge d \phi_A \wedge d \phi_B \wedge d \psi_A \\
    - e_A^2 \psi_A d r \wedge d \theta_A \wedge d \theta_B \wedge d \varphi \wedge d \phi_A \wedge d \phi_B \wedge d \psi_B \\
    + r \frac{(e_A^2+e_B^2)}{2} d \theta_A \wedge d \theta_B \wedge d \varphi \wedge d \phi_A \wedge d \phi_B \wedge d \psi_A \wedge d \psi_B ).
\end{aligned}
\end{equation}

We then choose the configuration containing strings at $\psi_A = e_B^2/\#$ and $\psi_B = 
e_A^2/\#$, where $\#$ 
stands for a normalization, since we know that we must have $\psi_{A,B}<4\pi$ 
(which we will ignore from now on, since we are not so much after an exact quantization
condition, but after the existence or not of various branes in the geometry). 
This specific configuration, as claimed before, is consistent with $\lim_{r \rightarrow \infty} \rho(r) = 0$,  
where $\rho(r)$ 
is the radial effective density obtained after integrating the compact part of the cycle, $\rho(r) =
 \int d \Omega \rho_1$. In fact, we have
\bea
    \rho(r) &=& \int_{\text{compact}} \underbrace{\star_{10}\left ( e^{-2 \phi} H \right ) - C_2 \wedge \star 
    (B \wedge F_3)}_{\rho_1} \cr
    & =& \frac{32 \eta}{e_A^4 e_B^4 r^3} (e_B^3 Q_A + e_A^3 Q_B) (e_B Q_B - e_A Q_A) \text{Vol}_{S_3}^2 
    dr\;,
\eea
which is finite after integrating from $r = r_+$ to $r = \infty$. We finally obtain the 
number of fundamental strings $N_1$,
\bea
    N_1 &=& \int \frac{32 \eta}{e_A^4 e_B^4 r^3} (e_B^3 Q_A + e_A^3 Q_B) (e_B Q_B 
    - e_AQ_A) \text{Vol}_{S_3}^2 dr \cr
    &=& \frac{16 \eta}{e_A^4 e_B^4 r_+^2} ( e_B^3 Q_A + e_A^3 Q_B) |(e_B Q_B - e_A Q_A)
     |\text{Vol}_{S_3}^2\;,
 \eea
where we have put the modulus since we  consider the number of branes as positive. We also see 
that $r_+$ acts as a cutoff for the integration. This is expected, since the background we consider
corresponds to an IR completion for the I-brane gauge theory. Therefore, we finally obtain 
$N_1$ and $N_3$ as (integer) constants and, moreover, we see that for a specific combination of 
$e_{A,B}$ and $Q_{A,B}$, such that $e_b Q_B = e_a Q_A$, both charges vanish, meaning that 
we have a configuration with no fundamental strings, nor D3-branes, but made up only of D5-branes.

Summarizing, the resulting configuration in the target space consists of two stacks of D5-branes 
wrapping distinct three-cycles, $(\psi_{A,B}, \theta_{A,B}, \phi_{A,B})$, each carrying positive charge. 
Additionally, there are D3-branes extended over $(x,r,\varphi)$ with flux over the two five-cycles 
($\psi_{A,B}, \theta_{A,B}, \phi_{A,B}, \phi_{B,A}, \theta_{B,A}$). We also have F1 strings, extended along 
the $ x$-direction, and located above $\psi_A$ and $\psi_B$, as represented in Table  \ref{table}. 
The resulting configuration in the target space breaks all the supersymmetry that was left after twisting 
the background I of \cite{Nunez2023} to obtain the background II we considered here, 
which is expected, since our TsT deformation does not correspond  to a susy-preserving deformation. 

\begin{table}[h!]
\caption{Brane configuration in the given coordinate system}
\centering
\[
\begin{array}{|c|cccccccccc|c|}
\hline
 & t & x & r & \varphi & \theta_a & \phi_a & \psi_a  & \theta_b & \phi_b & \psi_b \\
\hline
F1 & \circ & \circ & \times & \times & \times & \times & \times & \times & \times & \times \\
D3 & \circ & \circ & \circ & \circ & \times & \times & \times & \times & \times & \times\\
D5_1& \circ & \circ & \times & \circ & \circ & \circ & \circ & \times & \times & \times  \\
D5_2 & \circ & \circ & \times & \circ & \times & \times & \times & \circ & \circ & \circ  \\
\hline
\end{array}
\]
\label{table}
\end{table}

\subsubsection*{The dual gauge theory}

Following the general principle of the AdS/CFT correspondence \cite{Maldacena_1999} in its  
stronger version, and moreover from its extension to gauge/gravity duality, meaning the correspondence
between string theory in a given background and a quantum field theory defined on its boundary, 
we expect the existence of a gauge theory (defined holographically).

As seen in \cite{Itzhaki2006,Nunez2023,Barbosa2024},  the dual gauge theory to the I and II backgrounds
of \cite{Nunez2023} are supersymmetric and confining. Since the spacetime geometry corresponds to a BPS supergravity solution, the dual 
field theory is expected to be supersymmetric. Moreover, the linear dilaton—which diverges as 
$r \to \infty$—indicates a running coupling that increases with energy, a behavior typically associated 
with asymptotic freedom in the dual gauge theory, while \cite{Nunez2023}
showed that the smooth ending of the space in the IR (with cigar-like topology) does indeed lead to 
confinement, and that the gauge coupling, computed from the action of a probe 5-brane, behaves 
essentially like the dilaton. However, the asymptotic freedom is unlike the undeformed case of 
\cite{Nunez2023}, where the coupling was bounded. See Appendix A.4 for details of the calculation.

In summary, as argued in \cite{Nunez2023}, Background II is conjectured to be dual to a 
three-dimensional supersymmetric gauge theory exhibiting confinement-like properties.

Here we consider $T\bar T$ single-trace
deformations, defined by TsT transformations,  
of the so-called Background II of the $I$-brane system, a fibered geometry 
preserving 4 out of 32 possible supersymmetries. The ten-dimensional supergravity solution is 
characterized by a non-trivial $C_2$ Ramond–Ramond field and a linear dilaton profile $\phi(r)$, 
with seven of the ten dimensions compactified. The compact space consists of two fibered 3-spheres, 
$S_3$, and a compact $\varphi$ direction.

After performing a TsT transformation on the original supergravity background, thereby generating a 
new geometry, via the rules in \cite{Lunin_2005}, it is natural to expect—as commonly observed in 
the literature \cite{Frolov_2005, Delduc_2014}—that the resulting background is dual to a gauge theory 
which corresponds to a deformation of the original  field theory. Although identifying the precise nature 
of the deformation is generally a nontrivial task, valuable insights can be obtained by examining the 
structure of the deformed supergravity background.

Of course, the $T\bar T$ single-trace deformation 
(so by $\sim \sum _i T_i\bar T_i$) is actually only defined in the case of symmetric 
product spaces ($\sim \otimes_i {\cal M}_i$)
for the gauge theory, which is not the case here, and so we must find what the analog 
of that is in our case. 

Since the deformed geometry no longer preserves supersymmetry, unlike the undeformed one, it is 
reasonable to assume that the deformation is induced either by a non-supersymmetric operator, or 
by a modification of the underlying structure of the theory—such as promoting the product between 
fields to a noncommutative one, as done in \cite{Lunin_2005}—in such a way that the Lagrangian 
ceases to be invariant under supersymmetric transformations (a point to which we will return later).

To gain some insights into the resulting field theory, we can analyze specific limits of the deformed 
background. First, consider $r \rightarrow r_+$, recalling that the type I background is defined 
only for $r \in (r_+, r_\star)$. 
Given that $r_+ \sim \frac{Q_A}{e_A} = \sqrt{N_A} e_A$ (assuming the supersymmetry-preserving 
condition of the original background), a small value of $r_+$ corresponds to a small number of D5-branes. 
If we further assume that $\eta$ is not large, then \eqref{eq: final back} reduces to
\begin{equation}
\begin{aligned}
    ds_{st}^2 &\sim r_+ (-dt^2 + dx^2) + (\dots) \\
    B &\sim 0 \\
    \Phi &\sim \ln r_+.
\end{aligned}
\label{tstlimit}
\end{equation}

In this limit, the $C_2$-field  remains unchanged under the deformation. 
The resulting expressions in \eqref{tstlimit} coincide precisely with the undeformed background 
given in \eqref{eq:metric original}. 

As shown in \cite{Nunez2023}, the type I and type II backgrounds coincide in the limit of small charges $Q$, 
which implies that the TsT-deformed version of the $I$ background should also reproduce the same structure 
in this regime. Alternatively, one could consider the case of small $\eta$, since the relevant small parameter is 
actually $(\eta r_+)^2$.

According to the holographic dictionary, the radial coordinate in the gravity side corresponds to the 
energy scale in the dual gauge theory. The fact that the deformed and undeformed backgrounds 
coincide near small $r$ suggests that, in the infrared regime, both gauge theories share the same behavior. 
This indicates that the deformation driving the gauge theory away from the original model may be governed by 
an irrelevant operator. In fact, connections between TsT transformations and irrelevant deformations have 
already been observed in the literature—for instance, in the case of the "single-trace" $T\bar{T}$ deformations 
\cite{giveon2020tbartlst}, where the dual gauge theory is two-dimensional. Here we don't have a symmetric
product CFT, but it is natural to assume that the TsT transformation gives rise to a deformation of a similar
type  to the single-trace $T\bar T$ one. 
Generalizations of $T\bar T$ deformations to higher-dimensional gauge theories have also been 
suggested in recent works, but we will not be needing them here. 

For the analysis of the dual field theory, the charges we have obtained, (D5, D3, F1), should give us 
information about the result. A simplification arises if we consider the S-dual of the theory, which in the 
field theory just amounts to a description exchanging weak and strong coupling. The S-dual maps the 
charges to (NS5, D3, D1) charges, which were studied in the literature 
\cite{giveon2020tbartlst}, \cite{Aharony:1998ub}. The theory, initially in 1+1 dimensions, but effectively, 
as we said, extended to 2+1 dimensions, then still contains (as in \cite{Barbosa2024}) the 
gauge group (with Chern-Simons levels) $ SU(N_A)_{N_B} \times SU(N_B)_{N_A}$, 
with Yang-Mills and Chern-Simons terms, and fermion bifundamentals under both groups, but now 
also with Yang-Mills groups for $SU(N_1)\times SU(N_3)$, due to the D1 and D3 charges,
and still has a global Lorentz
$SO(1,1)$ symmetry, not broken by the $B_{tx}\propto \epsilon_{t,x}$. The R-symmetry, 
as seen from the solution (\ref{eq: final back}), is unchanged from the one in \cite{Barbosa2024}, 
namely $SO(4)\times SO(4)$, for the two $S^3$'s, since TsT acted only on $x,t$. The R-symmetry 
corresponds in field theory to a global symmetry rotating the scalar fields (and the fermions).

Supersymmetry is broken by the TsT deformation, but the deformation is harder to understand than the 
undeformed theory. For this, we turn to the Penrose limit, in which some amount of supersymmetry is restored, 
and the theory looks simpler. 


\subsection{Penrose limit of the TsT transformation ($T\bar T$ deformation)}

The Penrose limit theorem \cite{Penrose1976} asserts that, in the vicinity of any null geodesic, a 
spacetime solution of the Einstein equations can be approximated by the universal form
\begin{align}
ds^2 = dv \left( du + \alpha_1 dv + \sum_i \beta_j dy^i \right) + \sum_j \gamma_{ij} dy^i dy^j,
\end{align}
where $\alpha_1$, $\beta_j$, and $\gamma_{ij}$ are functions of the transverse coordinates $y^i$. 
After performing an appropriate rescaling by an overall scale $R$ and taking the limit when $R\rightarrow 0$, 
this metric reduces to a pp wave geometry.

To identify a suitable null geodesic in the background described by \eqref{eq: final back}, solving the 
geodesic equations of motion, we consider the path defined by
\begin{equation}\begin{aligned}
r &= r_+ \\
x &= 0, \\
\theta_A = \theta_B &= \frac{\pi}{2}, \\
\phi_A = \phi_B = \theta &= 0\;,
\end{aligned}\end{equation}
where $r_+$ is the larger of the two zeroes of $f_s(r)$.
We can easily check that this solves the (null) geodesic equations of motion.

Along this trajectory, the string Lagrangian simplifies to
\begin{align}
\mathcal{L} = \frac{2r_+ d\psi_A^2}{e_A^2} + \frac{2r_+ d\psi_B^2}{e_B^2} - \frac{r_+ dt^2}{1 
+\eta^2 r_+^2}.
\end{align}

Introducing new angular coordinates $\psi_A = e_A(\theta + \psi)$, $\psi_B = e_B(\theta - \psi)$, 
and setting $\theta = 0$, so that we have a motion on the equators of both $S^3$ spheres, but in 
opposite directions, which is of the type also considered in \cite{Barbosa2024}, the Lagrangian becomes
\begin{align}
\mathcal{L} = r_+ \left( 4 d\psi^2 - \frac{dt^2}{1 + \eta^2 r_+^2} \right).
\end{align}

This expression can be recast into light-cone coordinates via the transformation
\begin{equation}\begin{aligned}
\label{changeofcoordinates}
du &= 2 d\psi - \frac{dt}{\sqrt{1 + \eta^2 r_+^2}}, \\
dv &= 2 d\psi + \frac{dt}{\sqrt{1 + \eta^2 r_+^2}}\;,
\end{aligned}\end{equation}

leading to the simplified form

\begin{equation}\begin{aligned}
\mathcal{L} = r_+ du dv.
\end{aligned}\end{equation}

To fully define the null geodesic, we further impose  $u = 0$.

Next we obtain the Penrose limit of the metric around this null geodesic, 
and find that it satisfies the equation of motion up to a
constant conformal factor. First, we obtain the metric near the point
\be
\begin{aligned}
    \theta_{A,B} = \frac{\pi}{2} \\
    r = r_+.
\end{aligned}
\ee
Taking care of the divergent factors involving $f_s(r)^{-1}$, and as usual rescaling the metric by a scale 
$L^2$, we get 
\begin{equation}
\begin{aligned}
   L^{-2} ds^2 ={}& -\frac{r_+}{1+\eta^2 r_+^2} dt^2 + \frac{r_+}{1 +
    \eta^2 r_+^2} dx^2 + d \varphi^2 
    (r-r_+) r_+ f_s'(r_+) + \frac{4 dr^2}{(r-r_+) r_+ f_s'(r_+)} \\
    &+ \frac{2 r_+}{e_A^2} \Big[ \big( \cos \psi_A\, d \phi_A \left( 1 - \tfrac{1}{2} (\theta_A - \tfrac{\pi}{2})^2 
    \right) - d \theta_A \sin \psi_A \big)^2 \\
    &\quad + \big( \cos \psi_A\, d \theta_A + d \phi_A \left( 1 - \tfrac{1}{2} (\theta_A - \tfrac{\pi}{2})^2 \right) 
    \sin \psi_A \big)^2 \\
    &\quad + \big( - (\theta_A - \tfrac{\pi}{2}) d \phi_A + d \psi_A - e_A Q_A\, d \varphi\, (r - r_+)\, 
    \zeta'(r_+)^2 \big)^2 \Big] \\
    &+ \frac{2 r_+}{e_B^2} \Big[ \big( \cos \psi_B\, d \phi_B \left( 1 - \tfrac{1}{2} (\theta_B - 
    \tfrac{\pi}{2})^2 \right) - d \theta_B \sin \psi_B \big)^2 \\
    &\quad + \big( \cos \psi_B\, d \theta_B + d \phi_B \left( 1 - \tfrac{1}{2} (\theta_B - 
    \tfrac{\pi}{2})^2 \right) \sin \psi_B \big)^2 \\
    &\quad + \big( - (\theta_B - \tfrac{\pi}{2}) d \phi_B + d \psi_B - e_B Q_B\, d \varphi\, (r - r_+)\, 
    \zeta'(r_+)^2 \big)^2 \Big].
\end{aligned}
\end{equation}

We then change coordinates to
\be
\begin{aligned}
\label{3}
\varphi=\varphi, \;   \phi_{A,B} = \frac{\Phi_{A,B}}{L} , \;   \theta_{A,B} = \frac{\pi}{2} + \frac{\Theta_{A,B}}{L} ,\;    r = r_+ + \frac{\rho^2}{L^2} \\
    x = \frac{X}{L} , \;   \psi_{A,B} = e_{A,B} \left( \frac{\Theta}{L} \pm \psi \right) ,  \; 
    t = \sqrt{1+\eta^2 r_+^2} \left( \frac{v}{2} - \frac{u}{2L^2}\right)\;,
\end{aligned}
\ee
where $\psi = \frac{1}{4} ( v + \frac{u}{L^2})$.  Keeping only terms up to $\frac{1}{L^2}$ in $L^{-2}ds^2$
(that is, terms that remain finite in the $L\rightarrow\infty$ limit of the metric), 
the metric reduces to
\bea
 L^{-2}   ds^2 &=& \frac{r_+ du dv}{L^2} + \frac{2 r_+}{e_A^2 L^2} d \Phi_A^2 
 + \frac{2 r_+}{e_A^2 L^2} d \Theta_A^2 
    + \frac{2 r_+}{e_B^2 L^2} d \Phi_B^2 + \frac{2 r_+}{e_B^2 L^2} d \Theta_B^2 + \frac{4 r_+ d \Theta^2}{L^2} 
    \cr
   &&+ \frac{16 d \rho^2}{L^2 r_+ f_s'(r_+)} + \frac{\rho^2 r_+ f_s'(r_+) d \varphi^2}{L^2}  
   + \frac{r_+ dX^2}{L^2 ( 1 + \eta^2 r_+^2)} \cr
   &&   + dv \left ( \frac{r_+ \Theta_B}{e_B L^2} d \Phi_B 
    - \frac{r_+ \Theta_A}{e_A L^2} d \Phi_A + \frac{\rho^2 ( Q_B - Q_A) r_+ \zeta'(r_+)}{L^2} d \varphi \right ).
\eea

Following the same approach for the dilaton and flux fields, we obtain

\be
\begin{aligned}
    \phi = \frac{1}{2} \log \frac{r_+^2}{1  + \eta^2 r_+^2} \\
    B = \frac{\eta r_+^2}{2 L \sqrt{1 +\eta^2 r_+^2}} dv \wedge d X\\
    F_3 = \frac{1}{2 e_A e_B L^2} (e_B d \Phi_A \wedge d \Theta_A \wedge d v - e_A d \Phi_B \wedge d \Theta_B \wedge dv +\\ 2 e_A e_B \rho (Q_A - Q_B) \zeta'(r_+) d \rho \wedge d v \wedge d \varphi)\;,
\end{aligned}
\ee
with all the others fields being $0$. However, since the rescaling of the metric by $L^2$ amounts to an
initial rescaling of all coordinates by $L$, and then the redefinition of the Penrose theorem gives 
factors of $1/L$ for the transverse coordinates, ignoring the initial rescaling of the coordinates, we expect 
in the Penrose limit to have ($F_{(n+1)}=dC_{(n)}$, so $F_{vx_1...x_n}=n\d_{[x_n}C_{vx_1...x_{n-1}]}$, with 
$x_i$ the rescaled transverse coordinates)
\be
F_{vx_1...x_n}\sim \frac{1}{L^n}\Rightarrow C_{vx_1...x_{n-1}}\sim \frac{1}{L^n}
\ee
in order to have a finite flux in the limit. 

We see then that 
\be
F_{(5)}=(1+*_{10})B\wedge F_{(3)}\lsim dv\wedge dv=0
\ee
in the limit, $F_{(3)}$ is constant and finite, as we obtain correctly $\sim 1/L^2$, but 
$B\sim {\cal O}(1/L)\Rightarrow \infty$ (for finiteness, it should have been $\sim 1/L^2$). 
That seems like a problem, but that is a somewhat common occurrence (see for instance 
\cite{Itsios:2017nou}): the point is that this $B$ field is an infinite constant, and as such is a pure 
gauge, that can be gauged away, so dropped. Note that the next term in the $1/L$ expansion would be 
${\cal O}(1/L^3)$, which vanishes in the $L\rightarrow \infty $ limit, so in the end we can put 
$B=0$.\footnote{Note that the equations of motion for the resulting solution have been 
checked.}

We see that the Penrose limit of the TsT background yields an S-dual parallelizable pp-wave metric with a 
constant dilaton and a closed 3-form, $d F_3 = 0$. 

The background generated  is a solution of the supergravity type IIB equations of motion. In fact, we can see 
that by (setting $L=1$ in the above, or rather, since everything is now ${\cal O}(1/L^2)$,
correctly doing the $L\rightarrow\infty$ limit and)
doing the change of coordinates (from $\rho$ to $R$ and from $\varphi$ to $\Phi$)
\be\begin{aligned}
    \rho \rightarrow \frac{R L \sqrt{r_+ f_s'(r_+)}}{4} \\
    \varphi \rightarrow \frac{4}{r_+ f_s'(r_+)} \Phi\;,
\end{aligned}\ee
one obtains (focusing only on a part of the metric)
\bea
    ds^2 &=& r_+ du dv +d R^2 + R^2 d \Phi^2 \cr
&&    +dv \left ( \frac{r_+ \Theta_B}{e_B } d \Phi_B 
    - \frac{r_+ \Theta_A}{e_A } d \Phi_A + \frac{R^2  ( Q_B - Q_A) r_+ \zeta'(r_+)}{4} d \Phi \right ).
\eea

Changing coordinates again by
\begin{align}
    d R = \frac{z dz + y dy}{R} \\
    d \Phi = \frac{z dy - y dz}{R^2} \\
    v = -\frac{2V}{r_+} \;,
\end{align}
and rescaling the other coordinates making part of $\sum_{i,j} \delta_{ij} dy_i dy_j$, 
the specific part of the metric above reduces to 
\bea
    ds^2 &=&-2 du dV + dz^2 + dy^2  \cr
    &&+dV \left ( -\frac{\Theta_B}{e_b } d \Phi_B + \frac{ \Theta_A}{e_a } 
    d \Phi_A - \frac{(Q_B-Q_A) \zeta'(r_+)}{2} (z dy - y dz) \right ).
\eea
We then change $u$ to
\begin{align}
    u \rightarrow U = u - \frac{1}{2 e_B} \theta_B \Phi_B + \frac{1}{2 e_A} \Theta_A \Phi_A\;,
\end{align}
and then the whole metric finally is 
\bea
\label{metricbeforechange}
    ds^2 &=&
    -2 du dV + dz^2 + dy^2  +dV \left ( \frac{1}{e_B} (\Phi_B d \Theta_B - \Theta_B d \Phi_B) + \frac{ 1}{e_A } (\Theta_A d \Phi_A - \Phi_A d \Theta_A)\right.\cr
    &&\left. - \frac{(Q_B-Q_A) \zeta'(r_+)}{2} (z dy - y dz) \right ).
\eea


Doing the same coordinate transformation for  $F_3$, we get
\bea
     F_3 &=& \frac{1}{e_A e_B } \left(-\frac{1}{r_+} e_B d \Phi_A \wedge d \Theta_A \wedge d V + \frac{1}{r_+} 
     e_A d \Phi_B \wedge d \Theta_B \wedge dV\right.\cr
     &&\left. -  \frac{ 1}{2 r_+} e_A e_B (Q_A - Q_B) \zeta'(r_+) dz \wedge d V \wedge d y\right)\;,
\eea
which is obtained from
\bea
    C_2 &=& \frac{1}{2e_A r_+} dV \wedge ( \Theta_A d \Phi_A - \Phi_A d \Theta_A) - \frac{1}{2e_B r_+} dV 
    \wedge ( \Theta_B d \Phi_B - \Phi_B d \Theta_B )\cr
    && +\frac{ 1}{4 r_+} (Q_A - Q_B) \zeta'(r_+) dV \wedge (z dy - y dz)\;.
\eea



Since the background has a constant dilaton, and it is S-dual of the parallelizable pp wave background, we 
know that it is a supersymmetric solution preserving, if $Q_A = Q_B$ (zero coefficient of $dV dy$ and $dV
dz$ in the metric) and $e_A=e_B$ (equal contributions for the remaining transverse
coordinates in the metric), $24$ supersymmetries. Note that, since $e^2_{A,B}=8/N_{B,A}$ (from the 
background before TsT), $e_A=e_B$ means $N_A=N_B$. 

In this case, since, as shown in \cite{Nunez2023}, the original solution (before TsT), was supersymmetric
if $e_AQ_B=\pm e_B Q_A$, or $Q_B\sqrt{N_A}=\pm Q_A \sqrt{N_B}$, it means that in our case, the 
pp wave with 24 supersymmetries corresponds to an original solution, so undeformed field theory as well, 
that was supersymmetric, before TsT.

Moreover, we see that the condition $Q_A=Q_B$ makes $N_{D3}$ exactly $0$. This is interesting, 
though not so surprising: It is hard to find a complex configuration involving a large number of different 
branes that preserves supersymmetry, and therefore we should expect that we need to get rid of some 
of the branes in order to maintain something.

\subsubsection*{The string action and the operator-state map in gauge/gravity dual}

To obtain the string spectrum for the pp wave solution with metric \eqref{metricbeforechange} and $B=0$,
we must
substitute the background in the Polyakov string action. We see that we have a nontrivial $C_2$ and 
$F_3$, but that will affect only the fermionic action, which we will not derive here. We will only 
derive the bosonic spectrum, to be matched with a corresponding spectrum in the field theory.





However, before we study the spectrum, we are going to do another change of coordinates.  
From  \eqref{metricbeforechange}, we can do the change of coordinates analogous to the 
one done in \cite{Araujo_2017}, \footnote{$z=x+iy=e^{-iaV/2}\omega$}
\begin{align}
\label{rule}
    dV \left ( a (xdy - ydx) \right) + dx^2 + dy^2 \rightarrow dV^2 \frac{a^2}{4} |\omega|^2 
    + d \omega d \bar{\omega}.
\end{align}
 We therefore obtain, after some rescaling,
\bea
    ds^2 &=&    -2 du dV   - dV^2 \left (r_+ \frac{e_B^2}{16 } |\omega_1|^2 + r_+ \frac{ e_A^2}{16 }
    |\omega_2|^2 + \frac{(Q_B-Q_A)^2 \zeta'(r_+)^2}{16} |\omega_3|^2 \right ) \cr
    &&+ |d\omega_1|^2 + |d\omega_2|^2 + |d\omega_3|^2 + \dots \cr
        \Phi &=& \frac{1}{2} \log \frac{r_+^2}{1 + \eta^2 r_+^2}.
\eea

We see that the TsT deformation by $\eta$ really has only affected the constant dilaton $\Phi$, therefore 
the string coupling, or the YM coupling in the field theory dual, otherwise there is no difference.

The Polyakov action is
\begin{align}
    S = - \frac{1}{4 \pi \alpha'} \int d \sigma d \tau \left ( \gamma^{ab} \sqrt{\gamma} G_{\mu \nu} \partial_a 
    X^{\mu} \partial_b X^{\nu} + \epsilon^{ab} B_{\mu \nu} \partial_a X^{\mu} \partial_b X^{\nu} + 2 \pi \alpha' \Phi 
    \mathcal{R} \right ).
\end{align}
Since $\Phi$ is constant, its contribution to the action is a topological term (the Euler characteristic 
\cite{do2016differential}), and does not contribute to the analysis of the spectrum, so we can simply ignore it, 
and $B=0$.

The pp wave is S-dual to a parallelizable pp wave, meaning that instead of having a NS-NS $B_2$ field 
of the form in (\ref{pp wave original}), we have a R-R $C_2$ field of a similar form, while the pp wave 
metric is diagonal in the transverse coordinates. But that means that only the fermions will be 
affected by $C_2$, and we will not consider here their frequencies. The bosons have simple frequencies, 
calculated as in the standard ${\cal N}=4$ SYM and ABJM cases. 

Using the light-cone gauge $X^V = \tau$, the action reduces to (with $\gamma_{ab} \rightarrow \eta_{ab}$), 
\bea
    S &=& -\frac{1}{4 \pi \alpha'} \int d^2\sigma\left[\sum_{i \neq u,V} \eta^{ab} (\partial_a X^i 
    \partial_b X^j) G_{ij} - 
    \left ( r_+\frac{e_B^2}{16} |\omega_1|^2 +r_+ \frac{e_A^2}{16}|\omega_2|^2 \right.\right.\cr
    &&\left.\left.+ \frac{(Q_B-Q_A)^2 \zeta'(r_+)^2}{16} |\omega_3|^2 \right ) \right ]\;,
\eea
where we used $\epsilon^{\tau \sigma} = 1$. 
From the Lagrangian we see that
$\Theta$ and $X$ are free (massless) waves, while the others $6$ coordinates 
(counting that excludes $u,v$) are massive waves, with nontrivial frequencies. The equation of motion are
\begin{equation}
\begin{aligned}
\label{eom}
    \Box_2 X^\Theta = 0 \\
    \Box_2 X^x = 0 \\
    \Box_2 \omega_{1,\bar{1}} + r_+ \frac{e_B^2}{16} \omega_{1,\bar{1}} = 0  \\
    \Box_2 \omega_{2,\bar{2}} + r_+ \frac{e_A^2}{16} \omega_{2,\bar{2}} = 0  \\
    \Box_2 \omega_{1,\bar{1}} + \frac{(Q_B-Q_A)^2 \zeta'(r_+)^2}{16} \omega_{3,\bar{3}} = 0 .
\end{aligned}
\end{equation}

 Before proceeding we should notice that the equations of motion away from the light-cone gauge are, 
 denoting by $\varepsilon = \left ( r_+\frac{e_B^2}{16} |\omega_1|^2 +r_+ \frac{e_A^2}{16}|\omega_2|^2 
 + \frac{(Q_B-Q_A)^2 \zeta'(r_+)^2}{16} |\omega_3|^2 \right ) = c_j \omega_j \bar{\omega}_j$,
\begin{equation}
\begin{aligned}
\label{eomnongauge}
    \Box_2 V = 0 \\
    \Box_2 U + \partial^a V \partial_a \varepsilon = \partial^2 U + \partial^a V c_j (\bar{\omega}_j 
    \partial_a \omega_j + \omega_j \partial_a \bar{\omega}_j )  = 0 \\
    \partial^2 \bar{\omega} + (\partial V)^2 c_1 \bar{\omega}_1 = 0.
\end{aligned}
\end{equation}

The light-cone gauge condition is satisfied and the equation of motion of the $\omega$'s 
reduces to \eqref{eom} if  $U$ satisfies the second equation of \eqref{eomnongauge}, 
\begin{align}
    \Box_2 U + c_j \partial_{\tau} ( \omega_j \bar{\omega}_j ) = 0.
\end{align}

This equation is implicit in our formalism, and it has a solution, however it is not useful for the case here. 
In fact, as we saw, in the light-cone gauge terms with  $U$ in the Lagrangian disappear, so the spin-chain
Hamiltonian  will be the same. 

As usual, one changes the light-cone gauge to the more standard $\d_\tau u=\a' p^+$, 
where $p^+=p^V$, and considers the 
standard ansatz, consistent with the periodicity of closed strings, 
\be
\xi_i=\xi_i^0e ^{in_i\sigma-i\omega t}\;, i=1,...,8\;,
\ee
where $\xi_i$ stands for all transverse coordinates, $(X^\Theta, X^x,\omega_1,\omega_{\bar 1},
\omega_2,\omega_{\bar 2},\omega_3,\omega_{\bar 3})$. 

Then the frequencies of oscillation for the $\omega$ fluctuations are obtained from putting the 
ansatz with the rescaled light-cone gauge into \eqref{eom},
\be\begin{aligned}
\Omega_{\Theta,x}=\frac{n_{\Theta,x}}{\a' p^+}\\
    \Omega_{1,\bar{1}} = \frac{e_B}{4}\sqrt{r_+} \sqrt{1+\frac{n^2_{1,\bar{1}}}{\a' p^+}} \\
    \Omega_{2,\bar{2}} = \frac{e_A}{4} \sqrt{r_+} \sqrt{1+\frac{n^2_{2,\bar{2}}}{\a' p^+}}\\
    \Omega_{3,\bar{3}} = \frac{|(Q_B-Q_A) \zeta'(r_+)|}{4}\sqrt{1+\frac{n^2_{3,\bar{3}}}{\a' p^+}}.
\end{aligned}\ee

These simple solutions give a map between states for the string on the pp wave, created by 
$a^{\dagger}_i(\Omega_j)$ operators acting on the vacuum, and operators on a sector of the 
deformed gauge theory, that we will describe in the last subsection. Since we are talking about 
closed strings, the Hamiltonian of this system is simply
\begin{align}
    H = \sum_{i=1}^8 \Omega_{i} N_i\;,
\end{align}
where the sum is over  $i=\Theta,x,1,\bar{1},2,\bar 2, 3, \bar 3 $.

As in the case in \cite{Barbosa2024}, the only case that is easy to understand in field theory
 is the supersymmetric
case, and moreover at $n_i=0$ (so only the states with no string momenta). We see that at $e_A=e_B$ and 
$Q_A=Q_B$ we have 4 vanishing frequencies, and 4 equal frequencies (which can be rescaled to 1 by 
a conformal factor).

\subsubsection{The operator-state mapping from string oscillators to the gauge theory}

The momenta $p_{u}$ and $p_v$ become, in gauge theory notation,
\begin{equation}
    \begin{aligned}
        p^-\equiv p^u = - p_V = \frac{\sqrt{1+ \eta^2 r_+^2}}{2} \Delta - \frac{J_{\psi}}{4} \\
        p^+\equiv p^V = - p_u = -\frac{ \frac{\sqrt{1+ 
        \eta^2 r_+^2}}{2} \Delta +\frac{J_{\psi}}{4}}{L^2}\;,
    \end{aligned}
\end{equation}
where  $\Delta,J$ are eigenvalues of the energy $H$, corresponding to scaling dimension, 
and momentum $p_{\psi} =\frac{1}{2}( e_A p_{\psi_A} + e_B p_{\psi_B})$, corresponding to 
a $U(1)$ (global) R-charge, respectively. 
The negative sign here is a result of the choice in doing the change of coordinates in 
\eqref{changeofcoordinates}, and does not affect the results. 
Therefore, the Penrose limit corresponds to a specific sector of the gauge theory in which the surviving 
operators (à la BMN \cite{Berenstein_2002}) have large charge $J_{\psi}$, approximately 
equal to a large conformal dimension $\Delta$,
\begin{align}
    J_{\psi} \simeq 2 \sqrt{1 + (\eta r_+)^2} \Delta \propto  L^2\;.
\end{align}
Here  $J_{\psi} = \frac{1}{2} (e_A J_{\psi_A} + e_B J_{\psi_B})$,\footnote{Note that we have chosen the 
relative sign, which amounts to a choice of the relative positive R-charge in field theory, which we can 
always do.} and therefore we can have operators 
with large $J_{\psi_A}$ charge, large $J_{\psi_B}$ charge, or a mixture of both. 

The supersymmetric case, which we are interested in, corresponds to $e_A=e_B$ (and $Q_A=Q_B$), 
so we have $J_\psi=e_A/2(J_{\psi_A}+J_{\psi_B})$. The supersymmetric (with 24 supercharges) 
pp wave has $U(2)\times O(4)$ R-symmetry, as shown in the original parallelizable pp wave paper
\cite{Sadri:2003bk}.  

 As usual (see \cite{Berenstein_2002}), we start by mapping the vacuum states of string with a 
 combination of operators on the gauge side. The mapping consists in relating the coordinates of the 
 supergravity solution with fields in the gauge side. Coordinates perpendicular to the worldvolume 
 of the branes are scalars in the field side,  an effective description of fluctuations of those coordinates, 
 see \cite{Polchinski_1995}. 

In this Penrose limit, the 8 scalars transverse to the wave pair up into 
4 complex ones, $(Z,\bar Z)$ and $(W,\bar W)$, with $U(2)$ acting on $(Z,W)$ and $(W_2,\bar W_2,
W_3,\bar W_3)$ acted upon by $SO(4)$. These will give 6 oscillators (and 2 coordinates extra, one 
for the radial coordinate, and one for the angle in which the null geodesic moves, just like in the 
${\cal N}=4$ SYM case -where one has 4 oscillators plus 2 extra coordinates-), to which we must 
add the covariant derivatives in the $(1+1)-$dimensional field theory directions, $D_i$, $i=1,2$.  
We define then, formally,
\be
Z=\phi^2_{(1)}+i\phi^6_{(2)}\;,\;\; W=\phi^3_{(1)}+i\phi^7_{(2)}\;,\;\;
W_2=\phi^4_{(1)}+i\phi^5_{(1)}\;,\;\;
W_3=\phi^8_{(2)}+i\phi^9_{(2)}\;,
\ee
where $(1)$ and $(2)$ refers to coordinates transverse to one or the other of the 5-branes. 

The naive correspondence between the string oscillators, coming from pp wave coordinates, and the 
field theory oscillators, would be 
\bea
&& (x_1,x_2)=(\rho,\varphi)\rightarrow D_i\;,\;\;\;
(x_7,x_8)=(x,\Theta)\Rightarrow W\;,\cr
&&(x_3,x_4)=(\theta_A,\phi_A)\rightarrow W_2\;,\;\;\;
(x_5,x_6)=(\theta_B,\phi_B)\rightarrow W_3\;,
\eea
but, as explained also in \cite{Barbosa2024}, this does not match, as expected, since the field theory is 
effectively $(2+1)-$dimensional. 

The radial coordinates 
\be
\bar u=\sqrt{\sum_i \phi^i_{(1)}\phi^i_{(1)}}\;,\;\;\;
\bar v=\sqrt{\sum_j \phi^j_{(2)}\phi^j_{(2)}}\;,
\ee
transverse to each of the D5-branes, are joined at $\bar u=\bar v=0$, obtaining the extra spatial coordinate 
called $w$, and replacing the original spacetime coordinate $x$, and still corresponding to the $x_7$ oscillator.

In 2+1 dimensions, a scalar has the {\em classical} 
(conformal) dimension 1/2, so we must define the unit of $J_\psi$ to be 1, so  that $E=p^-=0$ for it. 
But, of course, the implicit assumption is that the TsT deformation by a parameter $\eta$ is such that 
$\Delta_\eta \sqrt{1+(\eta r_+)^2}=1$, or $\Delta_\eta=1/\sqrt{1+(\eta r_+)^2}$. 
We redefine the zero of the energy by an $E_0=1$, and then we find that, in order to get the correct 
spectrum, we must rescale $H$ by $\mu=-2$ (this is the same procedure as was needed in 
\cite{Nastase2022,Barbosa2024a,Barbosa2024}). Summarizing,

\begin{center}
    \begin{tabular}{|c|c|c|c|c|c|c|c|c|}
    \hline
    field & $Z$ & $W$ & $\bar Z$ & $\bar W$ & $W_2,\bar W_2$ & $W_3,\bar W_3$ & $D_{x_i}$ & $D_w$\\
    \hline\hline
    $\Delta$ & 1/2 & 1/2 & 1/2 & 1/2 & 1/2 & 1/2 & 1 &1 \\
    \hline
    $J_\psi$ & -1 & -1 & 1 & 1 & 0 & 0 & 0 & 0 \\
    \hline
    $\Delta-J_\psi/2$ & 1 & 1 & 0 & 0 & 1/2 & 1/2 & 1 & 1\\
    \hline
    $H/\mu=\Delta-J_\psi/2-E_0$ & 0 & 0 & -1 & -1 & -1/2 & -1/2 & 0 & 0\\
    \hline
    oscillator & - & $x_8$ & - & - & $x_3,x_4$ & $x_5,x_6$ & $x_1,x_2$ & $x_7$\\
    \hline
    \end{tabular}
\end{center}

 Then the vacuum is acted upon by the $U(2)$ part of the R-symmetry, rotating $Z$ and $W$, so that 
 we have 
\be
|0_a;p^+\rangle =\frac{1}{\sqrt{J_\psi N^{J_\psi/2}}}\Tr[Z_a^{J_\psi}]\;,
\ee
where $Z_a=(Z,W)$, so either one can be inserted. Then we insert fields for oscillators 
from the table above inside the trace, in order to obtain the string oscillator states. 

Alternatively, more in line with what we obtained, we can define a unique vacuum (with no $a$ index), 
defined just by $Z^J$, and insertions of $W$ correspond to massless string oscillators $(x_8$)
insertions. In the case of no string momentum, $n_i=0$, there is no difference between the two interpretations, 
but at $n_i\neq 0$, the latter interpretation should be the one that matches. However, we will not 
attempt here to match the $n_i\neq 0$ case, and we will leave it for further work, as 
it is complicated, but will not change the $n_i=0$ matching for the (second) unique vacuum identification, 
as in, for instance, \cite{Araujo_2017,Itsios:2017nou} (also 
\cite{Nastase2022,Barbosa2024,Barbosa2024a}).
We just note here that the BMN operator analysis of the TsT deformed is case is essentially the same 
as in the undeformed case, with the deformation just acting on $\Delta_\eta$ and $J_\psi$ simultaneously, 
but otherwise keeping the solvability of the sector intact. The deformation should also modify the 
parameter $b$, which in gauge theory should be proportional to $g^2_{YM}\propto e^{\Phi}$, which 
we saw that gets modified by $1/\sqrt{1+\eta^2r_+^2}$, the same factor as in $\Delta_\eta$.
What we could learn in this case is exactly what one was able to learn in 
\cite{Nastase2022,Barbosa2024,Barbosa2024a} (and in \cite{Araujo_2017,Itsios:2017nou}).

\subsubsection{Spin chain from discretized string}

This works out as usual, but for completeness we will show it here. By discretizing the string Hamiltonian 
on the pp wave, we obtain a  Hamiltonian that in principle can be described in the gauge theory, 
via the action of a dilatation operator on field theory operators, via Feynman diagrams. 

We have the string Lagrangian density
\bea
4\pi \a'\mathcal{L} &=& \sum_{i \neq u,V} \eta^{ab} (\partial_a X^i \partial_b X^j) G_{ij} \cr
&&- \left ( r_+\frac{e_B^2}{16} 
    |\omega_1|^2 +r_+ \frac{e_A^2}{16}|\omega_2|^2 + \frac{(Q_B-Q_A)^2 \zeta'(r_+)^2}{16} |\omega_3|^2 
    \right ).
\eea
This reduces to 
\bea
4\pi\a' \mathcal{L} &=& \dot{X}_{\theta}^2 + \dot{X}_{x}^2 +  \sum_{l=1,2,3} 
\dot{\omega}_l \dot{\bar{\omega}}_l 
    -  X'^2_{\theta} - X'^2_{x} -  \sum_{l=1,2,3} \omega'_l \bar{\omega'}_l \cr
    &&- \left ( r_+\frac{e_B^2}{16} |\omega_1|^2 +r_+ \frac{e_A^2}{16}|\omega_2|^2 
    + \frac{(Q_B-Q_A)^2 \zeta'(r_+)^2}{16} |\omega_3|^2 \right ) .
\eea
The Lagrangian is $L = \int \mathcal{L}(\sigma) d\sigma$. We first discretize it to
\begin{align}
    L = \sum_i a^i \mathcal{L}(\sigma_i)\;,
\end{align}
where the spatial derivative is replaced by $X' \rightarrow \frac{(X_i - X_{i+1})}{a}$, where 
\bea
4\pi \a'{\cal L}(\sigma_i) &=& \dot{X_i}_{\theta}^2 + \dot{X_i}_{x}^2 +  \sum_{l=1,2,3} \dot{\omega_i}{}_l 
        \dot{\bar{\omega_i}}{}_l -  (X^{\theta}_i - X^{\theta}_{i+1})^2/a^2 - (X^x_i - X^x_{i+1})^2/a^2 \cr
        &&-  \sum_{l=1,3} (\omega^{l}_i - \omega^{l}_{i+1}) (\bar{\omega}^{l}_i - \bar{\omega}^{l}_{i+1})/a^2 \cr
    &&- \left ( r_+\frac{e_B^2}{16} |\omega^i_1|^2 +r_+ \frac{e_A^2}{16}|\omega^i_2|^2 + \frac{(Q_B-Q_A)^2 
    \zeta'(r_+)^2}{16} |\omega^i_3|^2 \right ) .
\eea

The Hamiltonian density is
\begin{align}
    \mathcal{H} = \dot{X_x}^2 + \dot{X_{\theta}}^2 + \sum_l \dot{\omega_l} \dot{\bar{\omega
    _l}} +   X'^x X'^x + X'^{\theta} X'^{\theta} + \sum_{l=1}^3\omega'_l \bar{\omega'}_l 
    + \sum_{l=1}^3 c_l^2 |\omega_l|^2.
\end{align}

The discretized Hamiltonian is then
\bea
    H &=& \sum_{i=1}^J \left ( \sum_{l=1}^3 c_l a_i^{\dagger l} a^l_i  +  \sum_{l=1}^3 \frac{ (\omega^{l}_i 
    - \omega^{l}_{i+1}) (\bar{\omega}^{l}_i - \bar{\omega}^{l}_{i+1})}{a^2}+  \frac{(X^{\theta}_i 
    - X^{\theta}_{i+1})^2}{a^2}\right.\cr
    &&\left. + \frac{(X^x_i - X^x_{i+1})^2}{a^2} +\dot X_x^2+\dot X_\theta^2 \right ) \;,
\eea
where the fields $\xi^l_i=$ ($\omega_i^l,\bar\omega_i^l$) 
are understood as $\simeq b(a_i^{\dagger l}+a_i^l)/\sqrt{2}$, and the relation between the length $L$ of the 
string, the number $J$ of sites and the intersite distance $a$ is 
\be
\frac{L}{a}=2\pi \a' p^+=\frac{J}{b\mu}.
\ee
If the same calculation in the ${\cal N}=4$ SYM would still hold (not entirely clear, since there is less susy,
so there could be, at the very least, a renormalization of the constant $b$), we would obtain the 
conjecture
\bea
    H &=&\sum_{1,\bar 1}\sum_{n_1,n_{\bar 1}}
    \frac{e_B}{4} \sqrt{r_+} \sqrt{1+4b^2 \sin^2 \left(\frac{\pi n_{1,\bar{1}}}{J}\right)} 
    A_{n_1,n_{\bar 1}}^{\dagger} A_{n_1,n_{\bar 1}} \cr
    && + \sum_{2,\bar 2}\sum_{n_2,n_{\bar 2}}\frac{e_A}{4} \sqrt{r_+} \sqrt{1+4b^2 \sin^2
    \left (\frac{\pi n_{2,\bar{2}}}{J}\right)} 
     A_{n_2,n_{\bar 2}}^{\dagger} A_{n_2,n_{\bar 2}}  \cr
    &&+ \sum_{3,\bar 3}\sum_{n_3,n_{\bar 3}}\left|\frac{(Q_B-Q_A) }{4} \zeta'(r_+)\right| 
    \sqrt{1+4b^2 \sin^2 \left(\frac{\pi n_{3,\bar{3}}}{J}\right)} 
    A_{n_3,n_{\bar 3}}^{\dagger} A_{n_3,n_{\bar 3}}\cr
    &&+\sum_{x,\theta}\sum_{n_x,n_\theta}2b\left|\sin\left(\frac{\pi n_{x,\theta}}{J}\right)\right|
    A_{n_x,n_\theta}^\dagger A_{n_x,n_\theta}.
\eea

But to check if this is true, we would need to consider string oscillator momenta, which we left for further work.

\section{Penrose limit plus TsT transformation ($T\bar T$ deformation?)}

Next, we consider the TsT transformation of the Penrose limit (the opposite order of the two actions). 

As has been checked in some articles, like \cite{Nastase2023}, there is generally a non-commutativity 
relation involving these transformation (the TsT deformation and the Penrose limit). However, this 
it not a rule, and it could happen that the end background is the same in both ways. That  
would imply interesting relations between sectors of a gauge model and its deformation.

The Penrose limit of the S-dual of $I$-brane type II background was evaluated in \cite{Barbosa2024}.
We start by reproducing these previous results, just reminding that the original background has only 
$F_3, \phi$ fields: First we approach the $r = r_+, \theta_{A,B} = \frac{\pi}{2}$ null-geodesic conditions, 
and then we do the change of coordinates as in \eqref{3}, with the only difference that the 
transformation for the time is actually is $t \propto \frac{v}{2} - \frac{u}{2 L^2}$. The metric goes to 
(after the $L\rightarrow \infty$ limit)
\bea
    ds^2 &=& r_+ du dv + \frac{2 r_+}{e_A^2} d \Phi_A^2 + \frac{2 r_+}{e_A^2} 
    d \Theta_A^2 + \frac{2 r_+}{e_B^2 } d \Phi_B^2 + \frac{2 r_+}{e_B^2 } d \Theta_B^2 
    + 4 r_+ d \Theta^2 \cr
    &&+ \frac{16 d \rho^2}{ r_+ f_s'(r_+)} + \rho^2 r_+ f_s'(r_+) d \varphi^2 \cr
&&    + r_+ dX^2   + dv \left ( \frac{r_+ \Theta_B}{e_B } d \Phi_B - \frac{r_+ \Theta_A}{e_A } 
d \Phi_A +\rho^2 ( Q_B - Q_A) r_+ \zeta'(r_+) d \varphi \right )\;,\label{penrosefirst}
\eea
and the dilaton and $F_3$ fields to
\bea
\label{penroseeq}
    \phi &=& \log r_+ \cr
    F_3 &=& \frac{1}{2 e_A e_B } \left( e_B d \Phi_A \wedge d \Theta_A \wedge d v 
    - e_A d \Phi_B \wedge d \Theta_B \wedge dv \right.\cr
    &&\left.+ 2 e_A e_B \rho (Q_A - Q_B) \zeta'(r_+) d \rho \wedge d v \wedge d \varphi\right).
\eea
This background preserves, if $Q_A=Q_B$ and $e_A=e_B$ (or $N_A=N_B$), 
$24$ supersymmetries. There is a remarkable similarity 
between this pp wave spacetime and the one obtained after the TsT transformation, but this is not 
so surprising, since the undeformed and the deformed background are exactly the same, apart from 
the $g_{tt},g_{xx}$ components and the RR/NSNS-fields, that vanish in the Penrose limit.

\subsection{TsT transformation ($T\bar T$ deformation?) after Penrose limit}

As explained in \cite{Frolov_2005}, the TsT transformation is done by picking two different coordinates 
that posses killing vectors $\partial_{x_i}$ of translations. Evaluating the background \eqref{penrosefirst} 
we notice we have several possibilities to do the Penrose Limit, with the most prominent being 
$u,v,\Theta,\varphi,X$. Moreover, we can evaluate the TsT transformation directly on the supergravity 
background, or we can try to deform the string world-sheet (the Polyakov action \cite{POLYAKOV1981207}) in
a specific way. 

\subsubsection{TsT transformation ($T\bar T$ deformation?) over $v,X$}

The TsT transformation that we did in the last section, before doing the Penrose limit, was over the 
coordinates $t$ and $x$, and since we want to compare what are the effects of doing a TsT after the 
Penrose limit, we naturally need to deform in these same coordinates. However, the metric we obtained 
\eqref{penrosefirst} is in terms of the more useful coordinates for the Penrose limit $u,v$, and since these 
coordinates are related to $t$ via $t = \frac{v}{2} - \frac{u}{2L^2}$, and for large $L$ the one that prevails is 
$v$,  we do a TsT over $v$ and $X\propto x$. However, since $v$ is now the timelike 
coordinate, we again Wick rotate it, do the TsT and then Wick rotate back.

After TsT transforming with parameter $\eta$, the metric \eqref{penrosefirst} gets a correction of the order 
of $\eta^2$, and therefore the full metric can be written as $\tilde{ds}^2 = ds^2 -
 \eta^2 \delta ds^2$, with 
$\delta ds^2$ being
\begin{equation}\begin{aligned}
\label{def}
    \delta ds^2 = -\frac{r_+^3}{4 e_A^2 e_B^2} ( -e_B \Theta_A d \Phi_A + e_A \Theta_B d \Phi_B 
    + e_A e_A du)^2\;,
\end{aligned}\end{equation}
and for the NS-NS fields, we have the intriguing result that the dilaton is constant, and the field strength 
$H=dB$ is constant,
\begin{equation}\begin{aligned} 
\phi = \log r_+ \\
    B = -\frac{\eta r_+^2}{2 e_A e_B} ( -e_B \Theta_A d \Phi_A + e_A \Theta_B d \Phi_B
      + e_A e_B d u)\wedge dX.
\end{aligned}\end{equation}

The compact form of the deformation of the metric, \eqref{def}, together with the field strength $H=dB$ 
being constant, may make one ask oneself if we can't transform the coordinates in the metric to reduce 
it to a parallelizable pp wave metric. However, we now show it is not possible to find a coordinate 
transformation that satisfies this. We do that by showing that invariant  objects in both metrics, like the 
Ricci scalar $R$, can't be adjusted to match, that is $R_{\text{parallelizable}} \neq R_{\text{deformed}}$, 
and therefore both metrics can't be connected by a transformation, and therefore 
they aren't the same spacetime.

First, from the metric above, we ignore the terms that are not relevant to our analysis, 
$\Theta, \rho, \varphi$ and $X$, since the terms of the metric involving them weren't affected 
by the deformation. Then we are left with, if $Q_A=Q_B$,
\begin{equation}
\begin{aligned}
\label{metricclaim}
    \tilde{ds}^2 &= r_+ \, du \, dv 
    + \frac{2 r_+}{e_A^2} \, d \Phi_A^2 
    + \frac{2 r_+}{e_A^2} \, d \Theta_A^2 
    + \frac{2 r_+}{e_B^2} \, d \Phi_B^2 
    + \frac{2 r_+}{e_B^2} \, d \Theta_B^2 \\
    &+ dv \left( \frac{r_+ \Theta_B}{e_B} \, d \Phi_B 
    - \frac{r_+ \Theta_A}{e_A} \, d \Phi_A  \right) - \eta^2 \delta ds^2.
\end{aligned}
\end{equation}
    
Using a Mathematica notebook, we have calculated the Ricci form, which takes the simple form
\begin{align}
\label{r1}
    R = -\frac{1}{32} \Big ( e_A^2 + e_B^2 \Big ) r_{+ } \eta^2.
\end{align}

The whole spacetime/metric is then of the form $\mathcal{M} = \mathbb{R}^4 \times M_6$, 
where $\mathbb{R}^4$ is Euclidean, and $M_6$ is the metric above. Since $M_6$ is curved,  
and $\mathbb{R}^4$ is flat, the whole spacetime has a total Ricci curvature $\neq 0$ 
(This wouldn't be true if, for example,  instead of $\mathbb{R}^4$ we had another spacetime that was
not flat. In fact, for example the AdS$_5 \times $S$^5$ metric has a zero net Ricci scalar). 

To prove that we cannot turn $M_6$ into a parallelizable pp wave metric form, we assume that we can, 
with the general parallelizable wave metric (here we use $10$ coordinates) \cite{Sadri:2003bk},
\begin{align}
\label{ppgener}
    ds^2 = -2 du dv - F(u, x_i) du^2 + 2 \sum_i A_i(u,x_j) du dx^i + \sum_{i} dx_i^2.
\end{align}

The Ricci scalar of this generalized pp wave metric was calculated in \cite{Duval1991}, and it was 
showed to be $R=0$, as a general property of any pp wave metric. Since \eqref{r1} is only $0$ for 
$\eta = 0$, which is the undeformed background, both Ricci scalars are different, and then the metric for a 
TsT transformation over $x,v$ (we are going to show later that we can obtain parallelizable wave, but for 
other pairs of coordinates) is not of parallelizable type.

We could also have guessed that the metric can't reduce to \eqref{ppgener} just by looking at the 
$\sum_i g_{ij} dx^i dx^j$ part of the metric, where the pp wave metric occurs when $g_{ij} = \delta_{ij}$, but 
that was not an invariant statement; the Ricci scalar calculation is. 
Otherwise, we could have at most a  "Bargmann metric" \cite{Duval1991}, a more general metric of the form
\begin{align}
    ds^2 = -2 du dv - F(u, x_i) du^2 + 2 \sum_i A_i(u,x_j) du dx^i + \sum_{ij} g_{ij} dx_i dx_j.
\end{align}

Returning to \eqref{penrosefirst} and \eqref{def}, we finally expand $\delta ds$, to obtain
\bea
        ds^2 &=& \frac{1}{4} \Bigg(
- \frac{r_{+} \left(-8 - \eta^2 r^2 \Theta_A^2 \right) d\Phi_A^2}{e_A^2}
- \frac{r_{+} \left(-8 - \eta^2 r^2 \Theta_B^2 \right) d\Phi_B^2}{e_B^2}
+ 16 r_{+} \, d\Theta^2\cr
&&+ \frac{8 r_{+} \, d\Theta_A^2}{e_A^2}
+ \frac{8 r_{+} \, d\Theta_B^2}{e_B^2}  
+ \eta^2 r^3 \, du^2
- \frac{2 r_{+} \Theta_B \, d\Phi_B \left(+\eta^2 r_{+}^2 du - 2 dv\right)}{e_B} 
+ 4 r_{+} \, du \, dv\cr
&&
+ \frac{2 r_{+} \Theta_A \, d\Phi_A \left(-\eta^2 r_{+}^2 \Theta_B \, d\Phi_B -
 e_B \eta^2 r_{+}^2 du 
- 2 e_B dv \right)}{e_A e_B} 
+ 4 r_{+} \, dX^2\cr
&&+ \frac{64 \, d\rho^2}{r_{+} f_s'} 
+ 4 \rho^2 r_{+} \, d\varphi^2 f_s'
\Bigg).
\eea

Changing coordinates by
\begin{equation}
\begin{aligned}
    \Theta \to \frac{\Theta}{2 \sqrt{r_{+}}} , \Phi_A \to \frac{e_A \, \Phi_A}{\sqrt{2 r_{+}}}, \Phi_B \to \frac{e_B \, 
    \Phi_B}{\sqrt{2 r_{+}}}, X \to \frac{X}{\sqrt{r_{+}}}, \rho \to \frac{\sqrt{r_{+} f_s'} \, \rho}{4}, \\
\varphi \to \frac{4 \, \varphi}{r_{+} f_s'}, \Theta_A \to \frac{e_A \, \Theta_A}{\sqrt{2 r_{+}}}, \Theta_B \to 
\frac{e_B\, \theta_B}{\sqrt{2 r_{+}}}, u \to \frac{2 (v-u)}{\eta  r_{+}^{3/2}}, v \to +
\eta \sqrt{r} \, u\;,
\end{aligned}
\end{equation}
we obtain
\bea
        ds^2& =& d\rho^2  + \rho^2 d\varphi^2 + dX^2  + d\Theta^2 \cr
&&         + \left(1 + \frac{1}{8} \eta^2 r_{+}^2 \Theta_A^2 \right)\, d\Phi_A^2
+ \left(1+ \frac{1}{8} \eta^2 r_{+}^2 \Theta_B^2 \right)\, d\Phi_B^2 \cr
&&- \frac{1}{4} \eta^2 r_{+}^2 
\Theta_A \Theta_B \, d\Phi_A \, d\Phi_B + d\Theta_A^2 
+ d\Theta_B^2 \cr
&&- dv^2  + du^2
- \frac{\eta r_{+} \Theta_A \, d\Phi_A \, dv}{\sqrt{2}}
+ \frac{\eta r_{+} \Theta_B \, d\Phi_B \, dv}{\sqrt{2}}.
\eea

We see that the third line of the metric mixes the coordinates $\Phi_{A,B}, \Theta_{A,B}$ in a nontrivial 
way, and we can't reduce to a trivial Euclidean part $\sum_i x_i^2$ on the metric. In fact, even the 
most general metric we cited here, the Bargmann metric, is hard to obtain: In this case we don't need  to 
worry about the part of the metric not involving $u,v$, but even these coordinates are mixed in a nontrivial way. 

Due to the nature of this metric it would be very hard to obtain the spectrum of the state directly from the 
usual quantization of the Polyakov action. Yet, some assumptions can be made, in order to make a 
connection between the supergravity solution and a deformed field theory. 
The Penrose limit selects a large charge operator sector of the field theory, and the TsT deformation 
should deform this sector.
The presence of a non-constant B-field indicates, in general, either a dipole-type, or a noncommutative
deformation of the field theory, so in this case, the product algebra of the large R-charge operators should
be so deformed. However, it would be very hard to define this, so we will leave it for further work.

\subsubsection{TsT transformation over $v,\varphi$ and $v,u$}

As explained in the beginning of the section, for completeness, and also to see whether there are 
deformations that be more easily analyzed, 
we consider also deforming in different pairs of isometric directions. In fact, we will find that some of the 
deformation pairs present similar behaviors from the field point of view, like being dual to 
irrelevant deformations, or involving dipole or noncommutative deformations.

\subsubsection*{TsT transformation over $(v,\varphi)$}



Starting with the pair $(v,\varphi)$, and to simplify the calculation, we can define
an $M$ matrix element \eqref{elementM} for the coordinates $(v, \varphi)$ (see Appendix \ref{appendicea2} for the definitions of the terms and their applications)
\begin{align}
\label{melement}
    M = \left ( 1 + \eta ( e_{v \varphi} - e_{\varphi v} \right ) - \eta^2 \det \left ( \begin{pmatrix}
    e_{vv} & e_{v \varphi} \\
    e_{\varphi v} & e_{\varphi \varphi} \end{pmatrix} \right )^{-1}\;,
\end{align}
where $e_{ij} = g_{ij}$, since $B=0$. Following the calculation from Appendix \ref{appendicea2}, we obtain the deformed metric,
\bea
    ds^2 &=& \frac{2r_+}{e_A^2} d \phi_A^2 
    + \frac{2 r_+}{e_B^2} d \phi_B^2 
    + \frac{2 r_+}{e_A^2} d \theta_A^2 
    + \frac{2 r_+}{e_B^2} d \theta_B^2 
    + 4r_+ \, d \theta^2    
    + r_+ \, dX^2 
    + \frac{16}{r_+ f_s'(r_+)} d\rho^2 \cr
    &&\quad + \frac{\eta^2 \rho^2 r_+^3 f_s'(r_+)}{Y} du^2 
    + \frac{4 \rho^2 r_+ f_s'(r_+)}{Y} d \varphi^2 \cr
    &&\quad + \frac{2 r_+Y + \eta^2 \rho^2 r_+^3 \theta_A^2 f_s'(r_+)}{e_a^2 Y} d \phi_A^2 
    + \frac{2 r_+ Y + \eta^2 \rho^2 r_+^3 \theta_B^2 f_s'(r_+)}{e_b^2 Y} d\phi_B^2 \cr
    &&\quad + dv \left( \frac{4 \rho^2 (Q_B - Q_A) r_+ \zeta'(r_+)}{Y} d \varphi 
    + \frac{4 r_+ \Theta_B}{e_b Y} d \Phi_B 
    - \frac{4 r_+ \Theta_A}{e_a Y} d \Phi_A \right) \cr
    &&\quad + du \left( \frac{-2 \eta^2 \rho^2 r_+^3 \Theta_A f_s'(r_+)}{e_a Y} d \Phi_A 
    - \frac{-2 \eta^2 \rho^2 r_+^3 \Theta_B f_s'(r_+)}{e_b Y} d \Phi_B \right) \cr
    &&\quad - 
    \frac{2 \eta^2 \rho^2 r_+^3 \Theta_A \Theta_B f_s'(r_+)}{e_a e_b Y} d \Phi_A d \Phi_B 
    + \frac{4 r_+}{Y} du \, dv\;,
\eea
where $Y = 4 + \eta^2 \rho^4 (Q_A - Q_B)^2 r_+^2 \zeta'(r_+)^2$, and there are also nonzero 
B and $\phi(\rho)$ fields present. It is important to emphasize that the dilaton 
$\phi$ becomes radial dependent, implying an energy scale dependence in the field theory. 
Again, the $C_2$ field remains the same, and we have a $C_4$ just like 
in the previous cases. The form of the metric suggest again that the TsT 
deformation is dual to an irrelevant operator, because as $\rho \rightarrow 0$, it reduces again to the  
undeformed pp wave metric. The general structure of the solution does not easily suggest a field 
theory interpretation, since it is a complex 
combination of coordinates with a not so obvious meaning from the dual side. However, we can try to evaluate 
the deformation of the special case $Q_A = Q_B$  of the pp wave, 
where (if also $e_A=e_B$) $24$  supersymmetries were 
preserved. Under this condition, the metric reduces to the simple form
\begin{equation}
\label{newmetric-noL}
\begin{aligned}
    ds^2 &= r_+ \, du \, dv 
    + \frac{2 r_+}{e_A^2} \, d \Phi_A^2 
    + \frac{2 r_+}{e_A^2} \, d \Theta_A^2 
    + \frac{2 r_+}{e_B^2} \, d \Phi_B^2 
    + \frac{2 r_+}{e_B^2} \, d \Theta_B^2 
    + 4 r_+ \, d \Theta^2 
    + \frac{16}{r_+ f_s'(r_+)} \, d \rho^2 \\
&    + \rho^2 r_+ f_s'(r_+) \, d \varphi^2 
     + r_+ \, dX^2
    + dv \left( \frac{r_+ \Theta_B}{e_B} \, d \Phi_B - \frac{r_+ \Theta_A}{e_A} \, d \Phi_A \right)\cr
    &  + \frac{\eta^2 \rho^2 r_+^3 f_s'(r_+)}{4 e_A^2 e_B^2}
    \left( e_A e_B \, du + e_A \Theta_B \, d \Phi_B - e_B \Theta_A \, d \Phi_A \right)^2 \;,
\end{aligned}
\end{equation}
where the last line represents the effect of the TsT  transformation in the metric. In this case it is easier 
to notice that as $\rho \rightarrow 0$,  the metric reduces to the initial case, but we will talk more about this 
later. Also, the dilaton field $\phi$ under this supersymmetry condition reduces to the initial dilaton value, 
$\phi = \log r_+$, a constant scalar. Finally, the B-field reduces to the simple form
\begin{equation}\begin{aligned}
    B = -
    \frac{\eta \rho^2 r_+^2 f_s'(r_+)}{2 e_A e_B } \left ( e_B \Theta_A d \Phi_A \wedge d \varphi 
    - e_A \Theta_B d \Phi_B \wedge d \varphi - e_A e_B du \wedge d \varphi \right )\;,
\end{aligned}\end{equation}
 with the new $F_5$ also having a $\rho^2$ dependence. Note that, even in the susy $e_A=e_B$ case, 
 the $H=dB$ field is not constant anymore, since then we have 
 \be
 B={\rm constant}\times \rho^2(\Theta_Ad\Phi_A-\Theta_B\Phi_B-e_A u)\wedge d\varphi.
 \ee

 Moreover, the metric \eqref{newmetric-noL} presents the same problem as the one calculated for $v, X$, 
 but even worse, since now the deformation is dependent of the coordinate $\rho$ as well. This makes the 
 analysis very difficult, since the Hamiltonian is not so easily diagonalized as it is for pp waves 
 (even less than for parallelizable pp waves), see \cite{Nastase2023}.

\subsubsection*{TsT transformation over $V, u$}

Moving on to a TsT transformation over $V,u$, the lightcone coordinate for pp wave, 
we first calculate the $M$ matrix element \eqref{elementM}, and find
\begin{align}
    M = (1+\eta^2)^{-1} \;,
\end{align}
obtaining for the deformed background
\bea
         ds^2 &=&    +\frac{2}{\eta^2 + 1} du dV   -
          \frac{1}{\eta^2+1} dV^2 \left ( \frac{e_B^2}{16 } r_+ |\omega_1|^2 
         + \frac{ e_A^2}{16 } r_+ |\omega_2|^2 + \frac{(Q_B-Q_A)^2 \zeta'(r_+)^2}{16} |\omega_3|^2 \right ) \cr
         &&+ |d\omega_1|^2 + |d\omega_2|^2 + |d\omega_3|^2 + \dots \cr
    B_2 &=& \frac{\eta}{\eta^2+1} d u \wedge dv \cr
    \Phi &=& \log r_+ - \frac{1}{2} \log ( 1 + \eta^2).
\eea

As before, we also find nontrivial $B$-field and dilaton $\phi$ fields. Moreover, we also have the 
$C_4$ field, which as we have seen is simply related to the $B$ and $C_2$ field, 
the last being intact under the TsT  transformation,
\bea
    C_4 &=& B_2 \wedge C_2 \cr
& =&   \frac{ -\eta V}{2 (1 +\eta^2 ) r_+^2 f_s'(r_+)}
 ( - e_B  f_s'(r_+) d \Phi_B \wedge d \Theta_B  + e_A  
f_s'(r_+) d \Phi_A \wedge d \Theta_A \cr
&& + 4 (Q_B-Q_A) \sqrt{r_+ f_s'(r_+)} \zeta'(r_+) \wedge d y \wedge dz ) \wedge du \wedge d V.
\eea

We notice that now there is a similarity between the Penrose limit after TsT transformation 
over $x,t$ and the current
case, the TsT transformation over $u,v$ after the Penrose limit, even though it is was not obvious
that they should be similar. 
In fact, the deformed metric obtained above is still a parallelizable pp wave type, and since the $B$ and 
 dilaton $\phi$ fields  are still constants, the analysis of the field theory dual sector (spin chain for operator
 sector) is simple, and similar to the cases done before. In fact, the BMN mapping is essentially the same 
 (up to some constant factors).

\subsubsection{General cases for TsT transformation}

Instead of investigating case by case, it is better to summarize the most important specific cases 
and find the differences and similarities between the TsT transformations over distinct pair of coordinates. 
Also, by using the example of the last subsection, where we saw that the effect of the TsT transformation
 in the general background is not so ilumminating, we focus here only on the susy condition, $Q_A=Q_B$ 
 (and $e_A=e_B$) for a clearer analysis. 

We first notice that doing a TsT transformation over $v$ and any other coordinates that is not $u$ produces 
a rather similar metric: Calling $\delta ds^2 = ds^2_{\text{TsT}} - ds^2$, we explicitly write the case for 
each other coordinate,
\begin{itemize}
    \item $(v,X)$ 
    \begin{equation}\begin{aligned}
        \delta ds^2 = +
        \frac{\eta^2 r_+^3}{4 e_A^2 e_B^2} ( -e_B \Theta_A d \Phi_A + e_A \Theta_B d \Phi_B + 
        e_A e_B du)^2.
    \end{aligned}\end{equation}
    \item $(v, \varphi)$
    \begin{equation}\begin{aligned}
        \delta ds^2 = +
        \frac{\eta^2 \rho^2 r_+^3 f_s'(r_+)}{4 e_A^2 e_B^2}( -e_B \Theta_A d \Phi_A + e_A 
        \Theta_B d \Phi_B + e_A e_B du)^2.
    \end{aligned}\end{equation}
    \item $(v, \Theta)$
    \begin{equation}\begin{aligned}
        \delta ds^2 = + 
        \frac{\eta^2 r_+^3}{e_A^2 e_B^2} ( -e_B \Theta_A d \Phi_A + e_A \Theta_B d \Phi_B 
        + e_A e_B du)^2.
    \end{aligned}\end{equation}    
\end{itemize}

Beside the remarkable similarity between the metrics, we notice that the $(v, \varphi)$ case is a 
radially dependent one, and moreover, as $\rho \rightarrow 0$ (which means $r \rightarrow r_+$) the 
deformed background reduces to the original one. This will be explored later, where we will show that 
all deformations over $\varphi$ have this structure. For now, we emphasize that this coincidence 
may suggest that there are some relations between the deformed gauge theories. 

In fact, since $\rho$ ranges over $(0, \infty)$, from the formulas above for the deformations, we
see that for some specific value of $\rho$ (which from the gauge side means for some 
energy) the $(v, \varphi)$ deformed gauge theory is equivalent to the deformed $(v,X)$ and the 
deformed $(v, \Theta)$, both of which are marginal-like deformations of the field theory. 
So, the $(v, \varphi)$ gauge field passes over the other two deformed fields theories, as the energy increases.

Moreover, the dilaton field is equal in the three cases above, and constant, equal to
\begin{equation}\begin{aligned}
    \phi = \log r_+.
\end{aligned}\end{equation}

The more notable difference between these deformations is how the B-field transforms,
\begin{itemize}
    \item $(v, X)$
    \begin{equation}\begin{aligned}
        B = -\frac{\eta r_+^2}{2 e_A e_B} ( -e_B \Theta_A d \Phi_A  + e_A \Theta_B d \Phi_B 
         + e_A e_B d u )\wedge dX .
    \end{aligned}\end{equation}
    \item $(v, \varphi)$
    \begin{equation}\begin{aligned}
        B = -\frac{\eta \rho^2 r_+^2}{2 e_A e_B} f_s'(r_+) ( -e_B \Theta_A d \Phi_A  + e_A 
        \Theta_B d \Phi_B  + e_A e_B d u)\wedge d\varphi .
    \end{aligned}\end{equation}    
    \item $(v, \Theta)$
    \begin{equation}\begin{aligned}
        B =- \frac{\eta r_+^2}{2 e_A e_B} ( -e_B \Theta_A d \Phi_A  + e_A \Theta_B d 
        \Phi_B + e_A e_B d u)\wedge d\Theta .
    \end{aligned}\end{equation}   
\end{itemize}
We see that one leg of $B$ is in the $X,\varphi, \Theta$, corresponding to the TsT coordinate.

This implies that, while the deformed dual gauge theory may be similar, there is 
still a difference between them, defined by the $B$ field, responsible for either a dipole-like deformation, 
or noncommutativity.
Indeed, Seiberg and Witten \cite{Seiberg_1999} have shown that configurations involving constant 
Kalb-Ramond field B, and open strings, can be interpreted as a non-commutative field theory 
involving operators and a modified product, the Moyal star product. In \cite{Lunin_2005} it was  
generalized to a non-constant (though still associative) $B$ field. 
Furthermore, we expect that TsT deformations and noncommutative field theory are related, 
as explained in \cite{Lunin_2005}, with star product equal to, in first order
\begin{equation}\begin{aligned}
    A(x) \cdot B(x)  \rightarrow  A(x) \star B(x) = A(x) \cdot B(x) + \frac{i}{2} \theta^{ij} \partial_i A \partial_j B+...
\end{aligned}\end{equation}
The Moyal star product in the constant $B$, therefore constant $\theta$, case is 
\be
A(x)*B(x)=\lim_{x\rightarrow x'}A(x)e^{i\theta^{ij}\d_i\d'_j}B(x').
\ee

The noncommutativity $\theta^{ij}$, the open string metric $G_{ij}$ and the open string coupling 
constant $g_o$ are related to the closed string supergravity background by
\bea
\label{thetatrans}
    G_{ij} &=& -(2 \pi \alpha')^2 (B g^{-1} B)^{ij} \cr
    \theta^{ij} &=& (B^{-1})^{ij}\Rightarrow (G+2\pi \a'\theta)^{ij}=(g+2\pi \a' B)^{-1}_{ij}\cr
    g_o &=& g_s \left(\frac{\det G}{\det g}\right)^{1/4}.
\eea

The closed string metrics of the three deformations above have the same structure. The
$(v,X)$ and $(v,\Theta)$ ones only differ by a constant factor, and there is a energy factor 
$\sim \rho^2f'_s(r_+)$ relating $(v, \phi)$ to the other two. The B field differs only by the direction of 
the noncommutativity,  but the general form (since $|B|^2$ are also essentially the same) is the same, 
again with the $(v,\varphi)$ having {\em the same} $\rho^2 f'_s(r+)$ dependence. In the open string metric 
and $\theta^{ij}$, the latter will translate into $1/ \rho^2f'_s(r_+)$, but that is as expected.

Moving on, there is another interesting relation among the TsT deformations, when the pair of coordinates 
involves the $\varphi$ one. As we saw before, the TsT over $(v, \varphi)$ generates a deformation in the 
gauge side analogous to what would be the deformation done by an irrelevant operator. We now show 
that this is not a coincidence: any TsT over $\varphi$ generates dual gauge theories deformed by an
irrelevant operator (since we already studied $(v, \varphi)$, we will ignore it here),
\footnote{Note, however, that since in these cases there was no temporal coordinate ($v$) 
involved, there was no need to make any Wick rotation. Indeed, there are no apparent singularities 
appearing.}
\begin{itemize}
    \item $(u, \varphi)$
    \begin{equation}
    \begin{aligned}
        \delta ds^2 = -\frac{\eta^2 r_+^3 f_s'(r_+)}{4} \rho^2 d V^2 \\
        \phi = \log r_+ \\
        B = -\frac{\eta \rho^2 r_+^2}{2} f_s'(r_+) d \varphi \wedge d v.
    \end{aligned}
    \end{equation}
    \item $(X, \varphi)$
    \begin{equation}
    \begin{aligned}
        \delta ds^2 = - \frac{\eta^2 r_+^3 f_s'(r_+)}{1 + \eta^2 \rho^2 r_+^2 f_s'(r_+)} \rho^2 ( dX^2 
        + \rho^2 f_s'(r_+) d \varphi^2) \\
        \phi = \log r_+ - \log \sqrt{1 + \eta^2 r_+^2 f_s'(r_+)\rho^2} \\
        B = -
        \frac{ \eta \rho^2 r_+^2}{1+ \eta^2 \rho^2 r_+^2 f_s'(r_+)} f_s'(r_+) d \varphi \wedge d X.
    \end{aligned}
    \end{equation}
    \item $(\Theta, \varphi)$
    \begin{equation}
    \begin{aligned}
        \delta ds^2 =- \frac{4 \eta^2 r_+^3 f_s'(r_+)}{1 +
         4 \eta^2 \rho^2 r_+^2 f_s'(r_+)} \rho^2 
        (4 d \Theta^2 + f_s'(r_+) \rho^2 d \varphi^2 )\\
        \phi = \log r_+ - \log \sqrt{1 + 4 \eta^2 r_+^2 f_s'(r_+) \rho^2}\\
        B = -
        \frac{4 \eta \rho^2 r_+^2 f_s'(r_+)}{1+
         4 \eta^2 \rho^2 r_+^2 f_s'(r_+)} d \varphi \wedge d \Theta.
    \end{aligned}
    \end{equation}
\end{itemize}

Therefore, we see that every metric deformed by a TsT transformation that involves $\varphi$ 
has a structure of the type $\delta ds^2 \sim \rho^2 \dots$, and reproduces in the 
supergravity side the behaviour of {\em an irrelevant operator} in the gauge side. 

Besides that, 
we see two interesting facts regarding these deformations: while the $(\Theta, \varphi)$ and 
$(X, \varphi)$ cases vanish for $\rho \rightarrow 0$, and the dilaton goes to a constant, 
for $\rho \rightarrow \infty$ the B field goes to a constant (and therefore, can be gauged away; though 
one could still have noncommutativity, in principle), 
and the dilaton goes to $\phi \rightarrow - \log \rho$, which means that the coupling constant 
$g_s = e^{\phi} \sim \frac{1}{\rho}\rightarrow 0$, meaning that there is no noncommutativity,
and therefore the deformed gauge side is free, and perturbative 
at high energies, with operators interacting via the usual product. This is a remarkable result, 
since this suggests that we have an {\em asymptotically free gauge theory}, at least in some 
large R-charge sector. It is, however, true that the dilaton alone does not define the 
gauge coupling in these cases, one has to consider a probe DBI brane action in the background for 
that, as in \cite{Nunez2023}. We do this in Appendix A.4, where we find that the behaviour of the 
dilaton is qualitatively the same as of the gauge coupling, at least in the case of the TsT transformed 
I-brane background, so the conclusion stands.

The $(u, \varphi)$ case is different form the others; while its gauge dual also can be described as 
an irrelevant operator, there is no asymptotic freedom, and the B field increase as we go to higher energies. 
However, there is a still more striking observation about the metric of this deformed model: it is still a 
parallelizable pp wave! In fact, we will analyze it in more detail next.

\subsubsection{TsT transformation over $(u, \varphi)$}

The metric is (without restricting to the susy preserving case)
\bea
    ds^2 &=& r_+ \, du \, dv 
    + \frac{2 r_+}{e_A^2} d \Phi_A^2 
    + \frac{2 r_+}{e_A^2} d \Theta_A^2 
    + \frac{2 r_+}{e_B^2} d \Phi_B^2 
    + \frac{2 r_+}{e_B^2} d \Theta_B^2 
    + 4 r_+ \, d \Theta^2  \cr
    &&\quad + \frac{16}{r_+ f_s'(r_+)} d \rho^2 
    + \rho^2 r_+ f_s'(r_+) \, d \varphi^2 
    + r_+ \, dX^2  \cr
    &&\quad + dv \left( \frac{r_+ \Theta_B}{e_B} d \Phi_B 
    - \frac{r_+ \Theta_A}{e_A} d \Phi_A 
    + \rho^2 (Q_B - Q_A) r_+ \zeta'(r_+) d \varphi \right) \cr
    &&\quad - \frac{\eta^2 r_+^3 f_s'(r_+)}{4} \rho^2 \, dv^2.
\eea

We do the same change of coordinates as before,
\begin{equation}\begin{aligned}
    \rho \rightarrow R \frac{ \sqrt{r_+ f'_s(r_+)}}{4} \\
    \Phi \rightarrow \Phi \frac{4}{r_+ f'_s(r_+)} \\
    \Phi_{A,B} \rightarrow \Phi_{A,B} \frac{e_{A,B} }{\sqrt{2 r_+}} \\
    \Theta_{A,B} \rightarrow \Theta_{A,B} \frac{e_{A,B} }{\sqrt{2 r_+}} \;,
\end{aligned}\end{equation}
obtaining
\begin{equation}
\begin{aligned}
    ds^2 &= r_+ \, du \, dv 
    + d \Phi_A^2 + d \Theta_A^2 + d \Phi_B^2 + d \Theta_B^2 
    + 4 r_+ \, d \Theta^2 
    + d R^2 + R^2 d \Theta^2 \\
    &\quad + r_+ \, dX^2 \\
    &\quad + dv \left( \frac{e_B \Theta_B}{2} d \Phi_B 
    - \frac{e_A \Theta_A}{2} d \Phi_A 
    + \frac{R^2 (Q_B - Q_A) r_+ \zeta'(r_+)}{4} d \Phi \right) \\
    &\quad -\frac{\eta^2 R^2 r_+^4 f_s'(r_+)^2}{64} \, dv^2.
\end{aligned}
\end{equation}

Ignoring for a while the angular and $X$ terms, since them are equal, we  
focus on the expression involving $R, \Phi, v$,
\begin{equation}\begin{aligned}
    ds^2 = dR^2 + R^2 d \Phi^2 + dv \left ( \frac{R^2 (Q_B - Q_A) r_+ \zeta'(r_+)}{4} d\Phi \right ) 
    - \frac{ \eta^2 r_+^4 f_s'(r_+)^2}{64 L^4} R^2 dv^2.
\end{aligned}\end{equation}

We change coordinates to $R^2 = z^2 + y^2$, so
\begin{equation}\begin{aligned}
    ds^2 = dz^2 + dy^2 + dv \left ( \frac{(Q_B - Q_A) r_+ \zeta'(r_+)}{4} (z dy - y dz) \right ) 
    - \frac{\eta^2 r_+^4 f_s'(r_+)^2}{64 L^4} (z^2 + y^2 ) dv^2.
\end{aligned}\end{equation}

Then, defining $z_1 = z + i y$, we get
\begin{equation}\begin{aligned}
    ds^2 = d z_1 d \bar{z_1} - \frac{i}{2} dv \left ( \frac{(Q_B - Q_A) r_+ \zeta'(r_+)}{4} (\bar{z_1} d z_1 - z_1 d \bar{z_1}) \right ) - \frac{ \eta^2 r_+^4 f_s'(r_+)^2}{64 L^4} z \bar{z} dv^2.
\end{aligned}\end{equation}

We next do the rescaling $z_1 \rightarrow \omega e^{-\frac{iv a}{2}}$, where 
$a = \frac{Q_B - Q_A}{4} r_+ \zeta'(r_+)$, and $v \rightarrow \frac{-2V}{r_+}$, so we get
\begin{equation}\begin{aligned}
    ds^2 = d\omega d \bar{\omega} - \left(\frac{a^2 }{r_+^2} +
    \frac{ \eta^2 r_+^2 f_s'(r_+)^2}{16} \right)
    |\omega|^2 dV^2.
\end{aligned}\end{equation}

This shows that we have a parallelizable metric, since it is just like the previous case
 for the Penrose limit after TsT transformation, but with a new term,
\bea
\label{deformedmetric}
    ds^2 &=&    -2 du dV   -  dV^2 \left [r_+ \frac{e_B^2}{16 } |\omega_1|^2 + r_+ 
    \frac{ e_A^2}{16 }|\omega_2|^2 +\left(\frac{\eta^2 r_+^2 f_s'(r_+)^2}{16} \right.\right.\cr
&& \left.\left.   + \frac{(Q_B-Q_A)^2 \zeta'(r_+)^2}{16}\right) |\omega_3|^2 \right]
    +|d\omega_1|^2 + |d\omega_2|^2 + |d\omega_3|^2 + \dots \cr
    B_2 &=& i  \eta r_+ f_s'(r_+) ( \omega_3 d \bar{\omega}_3 
    - \bar{\omega}_3 d \omega_3 ) \wedge d V\cr
    \Phi& =& \log r_+.
\eea

This is more interesting than the $(v,u)$ case, since we have a true physical modification, not just a 
rescaling of coordinates.
In fact, we could obtain a spin-chain Hamiltonian just as we did before, and the fact that the two 
Hamiltonians (gauge and strings) match before we do the TsT transformations means that the deformed 
Hamiltonian must be the same for both cases. 
As we can see from \eqref{deformedmetric}, the new term deforms  the mass term of the Klein-Gordon field 
in the string (Polyakov) action. Moreover, the B part of the action will not be trivial anymore, and introduces 
new terms in the action, of the type  $\sim \omega_3 \partial_{\sigma} \omega_3 
- \bar{\omega}_3 \partial_{\sigma} \omega_3$.\footnote{Note that $B$ is still Hermitian, $B=\bar B$, 
and so is the string Lagrangian,
due to the $i$ in front.}


For the string action, using the gauge $X^V = \tau$, the new terms add 
\bea
    S &\supset&  \int d\sigma d\tau \left[|\partial \omega_3|^2 - \left(  
    \frac{\eta^2 r_+^2 f_s'(r_+)^2}{16} 
    + \frac{(Q_B-Q_A)^2 \zeta'(r_+)^2}{16} \right) |\omega_3|^2 \right.\cr
    &&\left.+ \frac{i \eta r_+ f_s'(r_+)}{2 } ( \omega_3 
    \partial_{\sigma} \bar{\omega}_3 - \bar{\omega}_3 \partial_{\sigma} \omega_3)\right].
\eea

\section{Interpretation as ("single-trace") $T\bar{T}$ deformation}

In the $AdS_3/CFT_2$ case with NS-NS fields, corresponding to a symmetric product CFT, 
it was shown in \cite{giveon2020tbartlst} that a "single trace" deformation, by $\sum_i T_i\bar T_i$
(instead of the standard $\sum_i T_i\sum_j \bar T_j$) is dual to a deformed holographic dual, 
which later, in \cite{Araujo_2019}, it was shown to be a TsT transformation, and has very similar properties
to the standard $T\bar T$ deformation. Therefore, in \cite{Nastase2023}, it was proposed that in 
other other cases and in other dimensions, like for instance in $AdS_5/CFT_4$, (perhaps several) 
TsT transformations correspond to some "generalized single-trace" $T\bar T$ deformations, 
that have similar properties to the $T\bar T$ deformation. Here, as we saw, we are still in 3 dimensions
in gravity, with CFT defined along the $x,t$ coordinates, though an extra coordinate appears effectively. 

Therefore, even if the "single-trace" interpretation is not directly relevant anymore, we believe that the same 
properties that make it similar to the standard $T\bar T$ as in \cite{giveon2020tbartlst}, in particular solvability,
should be still valid. This is indeed what we see in the TsT followed by Penrose limit case, that we 
analyzed first, where we saw that we just have a modification of $\Delta_\eta$ and $J_\psi$, 
as well as $g^2_{YM}$, by the same factor $1/\sqrt{1+
\eta^2r_+^2}$, but otherwise the system is as 
solvable as before the TsT deformation. 

In the case of a TsT deformation after the Penrose limit, things are less simple, since we saw that 
in the most relevant case, of deformation over $(X,v)$, the resulting string Hamiltonian is hard to 
analyze, as are most of the other cases. The cases of TsT deformations over $(v,X)$, $(v,\Theta)$ and 
$(u,\varphi)$ hold most promise to preserve solvability in a simple way, but we have not yet fully analyzed
them.




\section{Conclusion}

In this paper we have analyzed the combination of TsT transformation and Penrose limit, and 
the possible interpretation of the TsT transformation as a "generalized single-trace" $T\bar T$ 
deformation, in the case of the fibered $I$-brane solutions of \cite{Nunez2023}.

We have found that in the case of the TsT transformation on $(t,x)$ followed by a Penrose limit, 
we obtain a simple deformation of the undeformed case, preserving solvability, so of the generalized
$T\bar T$ type. 

In the case of TsT transformations of the Penrose limit, we have first found that there is no 
commutativity of limits. For the TsT transformation on $(X,v)$, corresponding to $(t,x)$ in the limit, 
we have found that it is hard to analyze the result. We have also considered TsT transformations
in other directions. 

In particular,  for transformations in $(X,\varphi)$ and $(\Theta,\varphi)$, we have 
found that we have an asymptotically free (and IR nontrivial!) 
gauge theory, at least in the relevant large R-charge sector, and in the $(u,\varphi)$ case 
we have again a parallelizable pp wave, though a deformed one, with a nonzero B field. 

There are many issues left for further work. The analysis of the spin chain for Penrose limit after 
TsT transformation in the case of nontrivial string momenta (interactions on the gauge theory side) is one. 
It would also be very interesting to study better the asymptotically free cases of TsT deformations in 
$(X,\varphi)$ and $(\Theta,\varphi)$ of the Penrose limit, as well as the parallelizable pp wave case of 
$(u,\varphi)$ deformations. Finally, we have always restricted the analysis of field theory to the 
supersymmetric case, but it would be interesting to see if we can say anything in the absence of susy.

\section*{Acknowledgements}

The work of HN is supported in part by  CNPq grant 304583/2023-5 and FAPESP grant 2019/21281-4. 
HN would also like to thank the ICTP-SAIFR for their support through FAPESP grant 2016/01343-7. 
MB is supported by FAPESP grant 2022/05152-2, and LS is supported by FAPESP grant 
2023/13676-4.


\appendix
\setcounter{equation}{0} 
\renewcommand{\theequation}{\Alph{section}.\arabic{equation}} 

\section{Appendix}

\subsection{Type IIB supergravity solutions}
\label{appendicea1}

The conventions we use to check if a type IIB background is a solution 
are the same as in \cite{Itsios_2012}, so we simply reproduce the important equations here. 

The action is 
\bea
        S_{IIB} &=& \frac{1}{2\kappa^2} \int \sqrt{-g} \, e^{-2\phi} \left[ \left( R + 4(\partial \phi)^2 - H^2 \right) 
        - \frac{1}{12} \left( F_1^2 + \frac{F_3^2}{3!} + \frac{1}{2*5!} F_5^2 \right) \right] \cr
        &&- \frac{1}{2} C_4 \wedge H \wedge dC_2\;,
\eea
where, since we have already taken in accounting the numerical factors $n!$, 
$F_p^2 = F_{\mu_1 \dots \mu_p}^{\mu_1 \dots \mu_p}$.

The fields are defined as
\begin{align}
    H = d B \quad F_1 = d C_0 \quad F_3 = d C_2 - C_0 H \quad F_5 = d C_4 - H \wedge C_2 + \dots\;,
\end{align}
and the equations of motion are
\begin{align}
    R_{\mu\nu} + 2 D_{\mu} D_{\nu} \Phi - \frac{1}{4} H_{\mu\nu}^2 &= e^{2\Phi} \left[ \frac{1}{2} 
    (F_1^2)_{\mu\nu} + \frac{1}{4} (F_3^2)_{\mu\nu} + \frac{1}{96} (F_5)^2_{\mu\nu} - \frac{1}{48} \eta_{\mu\nu} 
    \left( (F_1^2) + (F_3^2) \right) \right]
\end{align}
for the metric. For the other fields,
\be
\begin{aligned}
    R + 4 D^2 \Phi - 4 (\partial \Phi)^2 - \frac{1}{12} H^2 &= 0  \\
    d \left( e^{2\Phi} \star H \right) - F_1 \wedge \star F_3 - F_3 \wedge F_5 &= 0 \\
    d \star F_1 + H \wedge \star F_3 &= 0 \\
    d \star F_3 + H \wedge F_5 &= 0 \\
    d \star F_5 - H \wedge F_3 &= 0 \\
    F_5 = \star F_5.
\end{aligned}
\ee

\subsection{TsT transformation rules}
\label{appendicea2}

The derivation of the resulting background after doing a TsT transformation is easier to do if one 
knows how the fields transform under T-duality, which is well-known in the literature. Regarding the 
NS-NS fields, the T-duality transformations are known as the Buscher's rules, and were derived in 
\cite{BUSCHER198759}; we reproduce them here.

Doing a T duality over a coordinate we call "1", the rules for the NSNS-fields are (in string frame)
\begin{equation}
\begin{aligned}
    \tilde{G}_{11} &= \frac{1}{G_{11}}, \quad \tilde{G}_{1i} = \frac{B_{1i}}{G_{11}}, \\
    \tilde{G}_{ij} &= G_{ij} - \frac{G_{1i} G_{1j} - B_{1i} B_{1j}}{G_{11}}, \\
    \tilde{B}_{1i} &= \frac{G_{1i}}{G_{11}}, \\
    \tilde{B}_{ij} &= B_{ij} + \frac{G_{1i} B_{1j} - B_{1i} G_{1j}}{G_{11}}, \\
    \tilde{\Phi} &= \Phi - \frac{1}{2} \log G_{11}.
\end{aligned}
\end{equation}

For the RR-fields \cite{Meessen_1999} the rules consist of mappings among fields 
in type IIB to type IIA or vice-versa, 
\begin{equation}
\begin{aligned}
    \tilde{C}^{(n)}_{\mu...\nu\alpha 1} &= C^{(n-1)}_{\mu\nu\alpha} - (n-1) \frac{C^{(n-1)}_{[\mu...\nu|1} 
    G_{|\alpha 1}}{G_{11}}, \\
    \tilde{C}^{(n)}_{\mu...\nu\a\beta} &= C^{(n+1)}_{\mu...\nu\a\b 1} + n C^{(n-1)}_{[\mu...\nu\alpha} 
    B_{\beta]1} + n(n-1) \frac{C^{(n-1)}_{[\mu...\nu|1} B_{|\alpha|1} G_{|\beta]1}}{G_{11}}.
\end{aligned}
\end{equation}

Then, the TsT transformation consists of performing the previous transformation, shifting along another (isometric) coordinate, and then applying the T-duality rules again. This is a possible approach, but it involves lengthy algebraic manipulations, so we decided to follow the general rules for a TsT transformation derived in \cite{Imeroni_2008}. There, the element $\mathcal{M}$ is defined as follows: T-dualize along $2$ and shift along $3$, then

\begin{equation}
\label{elementM}
    \mathcal{M} = \left\{ 1 - \gamma \left( e_{\alpha 2 \alpha 3} - e_{\alpha 3 \alpha 2} \right) 
    + \gamma^2 \det \left( \begin{matrix}
    e_{\alpha 2 \alpha 2} & e_{\alpha 2 \alpha 3} \\
    e_{\alpha 3 \alpha 2} & e_{\alpha 3 \alpha 3}
    \end{matrix} \right) \right\}^{-1}.
\end{equation}

Using this element, the NSNS fields transformations under the TsT are, defining $e = B + g$,
\bea
    \tilde{\epsilon}_{MN} &=& \mathcal{M} \left\{ \epsilon_{MN} - \gamma \left[ \det \left( \begin{matrix}
    e_{\alpha 2 \alpha 3} & e_{\alpha N} \\
    e_{M \alpha 3} & e_{MN}
    \end{matrix} \right) \right] - \det \left( \begin{matrix}
    e_{\alpha 3 \alpha 2} & e_{\alpha N} \\
    e_{M \alpha 2} & e_{MN}
    \end{matrix} \right) \right.\cr
    &&\left.+ \gamma^2 \det \left( \begin{matrix}
    e_{\alpha 3 \alpha 2} & e_{\alpha 2 \alpha 3} & e_{\alpha 2 N} \\
    e_{\alpha 2 \alpha 3} & e_{\alpha 2 \alpha 3} & e_{\alpha 3 N} \\
    e_{\alpha 3 \alpha 3} & e_{M \alpha 3} & e_{MN}
    \end{matrix} \right) \right\}
\eea
\begin{equation}
    e^{2 \tilde{\phi}} = M e^{2\phi} \;,
\end{equation}
and the RR-fields are
\bea
    F_1 &=& f_1 + \gamma \left[ f_3 + f_1 \wedge b \right]_{[\alpha^2] [\alpha^3]} \cr
    F_3 + F_1 \wedge B &=& f_3 + f_1 \wedge b + \gamma \left[ f_5 + f_3 \wedge b 
    + \frac{1}{2} f_1 \wedge b \wedge b \right]_{[\alpha^2] [\alpha^3]} \cr
    F_5 + F_3 \wedge B + \frac{1}{2} F_1 \wedge B \wedge B &=& f_5 + f_3 \wedge b 
    + \frac{1}{2} f_1 \wedge b \wedge b \cr
    &&+ \gamma \left[ f_7 + f_5 \wedge b + \frac{1}{2} f_3 \wedge b \wedge b 
   + \frac{1}{6} f_1 \wedge b \wedge b \wedge b \right]_{[\alpha^2] [\alpha^3]}\;,\cr
   &&
\eea
where the indices for the brackets simply means contraction. One observation is that the 
equations of motion and Bianchi identities of generalized supergravity, obtained when taking the 
more general Yang-Baxter transformations, that include TsT as a special case, are very simple when written 
in terms of Page forms associated to Page charges\cite{Araujo:2017enj}.

\subsection{Page charges}
\label{appendicea3}

The method to obtain the Page charge-density $\rho$ follows from \cite{ferko2025holographynullboundaries}, but here the analysis is simpler. The fields are $H_3, B_2, F_5, C_4, F_3, C_2$.

\subsubsection{D3-brane}

We know that the D3-brane charge integrates over $5$-forms, 
and the possible combinations resulting in $5$-forms are
\begin{align}
    H_3 \wedge B_2, F_5, F_3 \wedge C_2, B_2 \wedge F_3, H_3 \wedge C_2 + \star( \dots ).
\end{align}

But $B_2 \wedge H_3= 0$ here, so we are left with
\begin{align}
     F_5, B_2 \wedge F_3, H_3 \wedge C_2 + \star( \dots ).
\end{align}

The term $F_3 \wedge C_2$ is discarded in order to obtain a closed form (similar steps are done in \cite{ferko2025holographynullboundaries}, but with a different configuration). Since $F_5 \sim B \wedge F_3$ already, in the end we essentially have $F_5$ combined with products of $H_3, C_2$.

In fact, from the equations of motion,
\begin{align}
    d \star F_5 - H \wedge F_3 = 0.
\end{align}

Substituting $F_5 = (1+\star_{10})B\wedge F_{3}$ and using $dF_3 = 0$, we obtain finally
\begin{align}
    d \star (B \wedge F_3) = 0.
\end{align}

This is the conserved current we were seeking, and then
\begin{align}
    Q_3 = \int_{\text{cycle}} \star B \wedge F_3
\end{align}
is constant.

\subsubsection{F1 fundamental string}

An $F_1$ string sources 3-forms. To see this, we start from the equation of motion, 
$ d \left( e^{-2\Phi} \star H \right)
- F_3 \wedge F_5 = 0 $.

Since $dF_5 = H_3 \wedge F_3$, 
\begin{align}
    d (C_2 \wedge F_5 ) = F_3 \wedge F_5 + C_2 \wedge d F_5 = F_3 \wedge F_5 
    + H_3 \wedge C_2 \wedge F_3.
\end{align}

And since $d C_2 \wedge F_3 = C_2 \wedge d F_3 = 0$,
\begin{align}
    H_3 \wedge C_2 \wedge F_3 = d ( B \wedge C_2 \wedge F_3)\;,
\end{align}
so we obtain that
\begin{align}
    d ( C_2 \wedge F_5 - C_2 \wedge B \wedge F_3) = F_3 \wedge F_5.
\end{align}

Substituting our solution $F_5 = ( 1+\star) B \wedge F_3$ in the left-hand side,
\begin{align}
    d(C_2 \wedge \star (B \wedge F_3) ) = F_3 \wedge F_5\;,
\end{align}
we see the right-hand side of the above equality is precisely the term we have in the equation of motion for $H_3$, and therefore we have showed it is an exact term. Finally, substituing it in the fundamental string equation, we obtain
\begin{align}
     d \left( e^{-2\Phi} \star H - C_2 \wedge \star (B \wedge F_3) \right) = 0.
\end{align}

Thus the current $j =e^{-2\Phi} \star H - C_2 \wedge \star (B \wedge F_3)$ is closed, 
and the fundamental string charge is
\begin{align}
    Q_1 = \int_{\text{cycle}} e^{-2\Phi} \star H - C_2 \wedge \star (B \wedge F_3).
\end{align}

\subsection{Gauge coupling calculation}

Here we calculate the gauge coupling for the TsT transformed I-brane background, by considering a 
probe D5-brane action in the background, just like in the original I-brane case in \cite{Nunez2023}.

As the TsT is performed over $(t,x)$, the wrapped cycles of the deformed background are exactly 
the same as for the undeformed one, i.e $\phi, \tilde{\Omega}_3, \hat{\Omega}_3$.
We embed the probe D5-brane in the background, and choose to wrap the $\tilde\Omega$ cycle, 
and be transverse to $\hat\Omega$ and $r$. 

The NSNS part of the DBI action is
\begin{equation}
    \begin{aligned}
        S_{\rm NS} = \int e^{-\phi} \mathrm{det} \sqrt{E + \alpha' F}.
    \end{aligned}
\end{equation}

Expanding it to first order in $\a'$,\footnote{Using  
\begin{equation}\begin{aligned}
    \sqrt{-\det (G+B + \alpha' F)} = \sqrt{-\det (E + \alpha' F)} = \sqrt{-\det E} \sqrt{\det (1 + \alpha' E^{-1} F)}\;,
\end{aligned}\end{equation}
where $E=G+B$, and 
\begin{equation}
    \begin{aligned}
        \sqrt{\det(1 + \alpha'\, c_1)} \approx 1 
+ \frac{1}{2} \alpha' \, \; \mathrm{tr}(c_1) 
+ \alpha'^2 \Bigg(  - \frac{1}{4} \; \mathrm{tr}(c_1^2) + \frac{1}{8} (\; \mathrm{tr}c_1)^2 \Bigg),
    \end{aligned}
\end{equation}
where $c_1 = E^{-1}F$.}
we obtain 
\bea
\label{eqNS}
    S_{NS} &=& - \mu \int d^6 \sigma \sqrt{-\det E} - \frac{\mu \alpha'}{2} \int \sqrt{-\det E} \; 
    \mathrm{tr}( E^{-1} F) \cr
    &&+ \frac{\mu \alpha'^2}{4} \int d^6 \sigma \sqrt{-\det E} \; \mathrm{tr} (F^2) + O(\alpha'^2).
\eea
We then follow the analysis of \cite{Filev:2007gb}, which considers an D7-brane embedding in a D3-brane
background. 
Thus, we first verify the consistency of the embedding. To do that, we need to take a look also on the 
WZ terms of the D5-brane. They are the terms linear on $C_i$  obtained by expanding 
$\int d^6 \sigma \sum_i C_i \exp \left(- F - \alpha' B \right )$, in the sign convention of \cite{Nunez2023},
\begin{equation}
    \begin{aligned}
        \int d^6 \sigma \Big [ -\frac{1}{3!} C_0 B \wedge B \wedge B - B \wedge C_4 + \frac{1}{2} B \wedge B \wedge C_2 + C_6 - \alpha' F \wedge C_4 + \frac{\alpha'^2}{2} F \wedge F \wedge C_2 \\
        + \alpha' B \wedge F \wedge C_2 - \frac{3}{3!} \alpha' C_0 F \wedge B \wedge B - \frac{3}{3!} \alpha'^2 C_0 F \wedge F \wedge B - \frac{\alpha'^3}{3!} C_0 F \wedge F \wedge F \Big ].
    \end{aligned}
\end{equation}
The constraint for the embedding comes from the first order in $F$ terms. Then, together with the fact 
that $B \wedge B = B \wedge C_4 = 0$, the full $\alpha'$ terms in the NSNS+RR action are
\begin{equation}
    \begin{aligned}
        \alpha' \mu_5 \int d^6 \sigma \sqrt{E} E^{[ab]} F_{[ab]} + \int d^6 \sigma \Big [ C_6 - \alpha' 
        F \wedge C_4 + \alpha' B \wedge F \wedge C_2  + O(\alpha'^2) \Big ],
    \end{aligned}
\end{equation}
and since $C_4 = B \wedge C_2$, the $\alpha'$ term vanishes, and we are left with
\begin{equation}
    \begin{aligned}
        \alpha' \mu_5 \int d^6 \sigma \sqrt{E} E^{[ab]} F_{[ab]} + \int d^6 \sigma C_6 + O(\alpha'^2).
    \end{aligned}
\end{equation}
The first term gives the constraint for $F$,
\begin{equation}
\label{equationE}
    \begin{aligned}
        \partial_a ( \sqrt{E} E^{[ab]} ) = 0.
    \end{aligned}
\end{equation}
Equation ~\eqref{equationE} is defined only for antisymmetric components of $E = G+B$, which come 
from the $B$ field, and the indices $(a,b)$ take value in $(t,x)$, but since neither the metric nor the 
$B$-field has components depending explicitly on these coordinates, the derivative vanishes and 
the left-hand side of~\eqref{equationE} is identically 0.
The second term in~\eqref{equationE} is a topological term, as it is proportional to the volume of the
probe brane. In fact, this term is also present before the TsT, and it gives the conserved charge of the 
D5 probe induced by the fields of the supergravity background. See~\cite{Nunez2023} for the explicit 
expression for $C_6$\footnote{In our case, however, the TsT introduces an overall factor of $\frac{1}{1 + 
\eta^2 r^2}$.}.

We are now ready to calculate the effective gauge coupling. 
The third term in~\eqref{eqNS} becomes
\begin{equation}
    \begin{aligned}
        -L_{\varphi} (4 \pi)^2 \left ( \frac{2}{e_A^2} \right )^{3/2} \frac{r^3}{1+r^2 \eta^2} e^{-\phi} \sqrt{f(r) 
        + 2 Q_B^2 \zeta(r)^2} \int dx dt \left ( \alpha'^2 \frac{F^2}{4} \right )\;,
    \end{aligned}
\end{equation}
where $L_{\varphi} =  \int_{\text{1 lap}} d\varphi$.
We must expand the $F^2$ term in the flat coordinates of the gauge theory, so we write
\begin{equation}
    \begin{aligned}
        F^2 = 2 F_{tx}^2 g^{tt} g^{xx} = \frac{2}{r^2} (1+r^2 \eta^2)^2 \eta^{tt} \eta^{xx} F_{tx}^2 = \frac{(1+r^2 \eta^2)^2}{r^2} f^2,
    \end{aligned}
\end{equation}
where $f$ is $2 F_{tx}^2 = F_{ij} F^{ij}$. Then,
\begin{equation}
    \begin{aligned}
        S = - \alpha'^2 L_{\varphi} (4 \pi)^2 \left ( \frac{2}{e_A^2} \right )^{3/2} r(1+r^2 \eta^2) e^{-\phi} 
        \sqrt{f(r) + 2 Q_B^2 \zeta(r)^2} \int dx dt  \frac{f^2}{4} = \\
        c_1 \mathrm{Vol}(t,x) - \alpha'^2 L_{\varphi} (4 \pi)^2 \left ( \frac{2}{e_A^2} \right )^{3/2} (
        1+r^2 \eta^2)^{\frac{3}{2}} \sqrt{f(r) + 2 Q_B^2 \zeta(r)^2} \int dx dt  \frac{f^2}{4}\;,
    \end{aligned}
\end{equation}
where in the last line we have substituted $e^{-\phi} = \frac{\sqrt{1 + \eta^2 r^2}}{r}$. From the above, we  
extract the coupling of the TsT-deformed gauge field, $g_{\rm YM}^{(\eta)}$, in terms of the 
underformed one, $g_{\rm YM,0}$.
\begin{equation}
    \begin{aligned}
        \frac{1}{g_{\rm YM}^{(\eta) 2}} = \alpha'^2 L_{\varphi} (4 \pi)^2 \left ( \frac{2}{e_A^2} \right )^{3/2} (1+r^2 
        \eta^2)^{\frac{3}{2}} \sqrt{f(r) + 2 Q_B^2 \zeta(r)^2} = \frac{1}{g_{\rm YM, 0}^2} 
        (1 + r^2 \eta^2)^{\frac{3}{2}}\;.
    \end{aligned}
\end{equation}

In the two asymptotic limits, we obtain
\begin{equation}
    \begin{aligned}
        \frac{1}{g_{\rm YM}^{(\eta) 2}} \; \Bigg\{
\begin{array}{ll}
\to 0 & \text{as } r \to r_+ \\[1mm]
\to \infty & \text{as } r \to \infty
\end{array}.
    \end{aligned}
\end{equation}

We see then that the behaviour of $g_{\rm YM}^{(\eta)}$ is qualitatively unchanged by the wrapping 
of the brane: we have a strong coupling limit near $r_+$, where it grows without limit, indicative of 
confinement, while 
in the UV regime, the coupling asymptotes to $0$, so asymptotically free. 
This is in contrast with the undeformed case, where 
the UV limit has a constant lower bound for $g_{\rm YM,0}$. 
Interestingly, the irrelevant deformation of the CFT by $\eta$ drives 
the gauge theory to a free UV point in the RG flow.

The fact that for the deformed theory we have an RG flow from non-perturbative strong coupling in the IR 
to an asymptotically free theory in the UV, that we have an IR cutoff introducing a length scale $r_+$, 
that is dual to a lower bound mass $m_g \sim r_+^{-1}$ representing a possible mass gap in the spectrum 
of the dual gauge theory, and the fact we have a supergravity solution with a domain-wall like behavior, are 
all suggesting  our dual field theory to be confining, though a final conclusion for this topic would require the 
calculation of the Wilson loop.


\bibliographystyle{utphys} 
\bibliography{TsT-penrose} 

\providecommand{\href}[2]{#2}\begingroup\raggedright\begin{thebibliography}{10}

\bibitem{Nunez2023}
C.~Nunez, M.~Oyarzo, and R.~Stuardo, ``{Confinement in (1 + 1) dimensions: a
  holographic perspective from I-branes},''
  \href{http://dx.doi.org/10.1007/JHEP09(2023)201}{{\em JHEP} {\bf 09} (2023)
  201}, \href{http://arxiv.org/abs/2307.04783}{{\tt arXiv:2307.04783
  [hep-th]}}.

\bibitem{Maldacena_1999}
J.~M. Maldacena, ``{The Large $N$ limit of superconformal field theories and
  supergravity},'' \href{http://dx.doi.org/10.4310/ATMP.1998.v2.n2.a1}{{\em
  Adv. Theor. Math. Phys.} {\bf 2} (1998)  231--252},
  \href{http://arxiv.org/abs/hep-th/9711200}{{\tt arXiv:hep-th/9711200}}.

\bibitem{Berenstein_2002}
D.~E. Berenstein, J.~M. Maldacena, and H.~S. Nastase, ``{Strings in flat space
  and pp waves from N=4 superYang-Mills},''
  \href{http://dx.doi.org/10.1088/1126-6708/2002/04/013}{{\em JHEP} {\bf 04}
  (2002)  013}, \href{http://arxiv.org/abs/hep-th/0202021}{{\tt
  arXiv:hep-th/0202021}}.

\bibitem{Araujo_2017}
T.~Araujo, G.~Itsios, H.~Nastase, and E.~{\'O}. Colg{\'a}in, ``{Penrose limits
  and spin chains in the GJV/CS-SYM duality},''
  \href{http://dx.doi.org/10.1007/JHEP12(2017)137}{{\em JHEP} {\bf 12} (2017)
  137}, \href{http://arxiv.org/abs/1706.02711}{{\tt arXiv:1706.02711
  [hep-th]}}.

\bibitem{Itsios:2017nou}
G.~Itsios, H.~Nastase, C.~N{\'u}{\~n}ez, K.~Sfetsos, and S.~Zacar{\'\i}as,
  ``{Penrose limits of Abelian and non-Abelian T-duals of $AdS_5\times S^5$ and
  their field theory duals},''
  \href{http://dx.doi.org/10.1007/JHEP01(2018)071}{{\em JHEP} {\bf 01} (2018)
  071}, \href{http://arxiv.org/abs/1711.09911}{{\tt arXiv:1711.09911
  [hep-th]}}.

\bibitem{Nastase2022}
H.~Nastase and M.~R. Barbosa, ``{Penrose limit of MNa solution and spin chains
  in three-dimensional field theories},''
  \href{http://dx.doi.org/10.1007/JHEP05(2022)181}{{\em JHEP} {\bf 05} (2022)
  181}, \href{http://arxiv.org/abs/2112.13802}{{\tt arXiv:2112.13802
  [hep-th]}}.

\bibitem{Barbosa2024a}
M.~R. Barbosa and H.~Nastase, ``{Penrose limit of T dual of Maldacena-Nastase
  solution and dual orbifold field theory},''
  \href{http://dx.doi.org/10.1103/PhysRevD.109.086025}{{\em Phys. Rev. D} {\bf
  109} (2024) no.~8, 086025}, \href{http://arxiv.org/abs/2307.15587}{{\tt
  arXiv:2307.15587 [hep-th]}}.

\bibitem{Barbosa2024}
M.~Barbosa, H.~Nastase, C.~Nunez, and R.~Stuardo, ``{Penrose limits of
  I-branes, twist-compactified D5-branes, and spin chains},''
  \href{http://dx.doi.org/10.1103/PhysRevD.110.046015}{{\em Phys. Rev. D} {\bf
  110} (2024) no.~4, 046015}, \href{http://arxiv.org/abs/2405.08767}{{\tt
  arXiv:2405.08767 [hep-th]}}.

\bibitem{Green1996}
M.~B. Green, J.~A. Harvey, and G.~W. Moore, ``{I-brane inflow and anomalous
  couplings on d-branes},''
  \href{http://dx.doi.org/10.1088/0264-9381/14/1/008}{{\em Class. Quant. Grav.}
  {\bf 14} (1997)  47--52}, \href{http://arxiv.org/abs/hep-th/9605033}{{\tt
  arXiv:hep-th/9605033}}.

\bibitem{Itzhaki2006}
N.~Itzhaki, D.~Kutasov, and N.~Seiberg, ``{I-brane dynamics},''
  \href{http://dx.doi.org/10.1088/1126-6708/2006/01/119}{{\em JHEP} {\bf 01}
  (2006)  119}, \href{http://arxiv.org/abs/hep-th/0508025}{{\tt
  arXiv:hep-th/0508025}}.

\bibitem{Zamolodchikov:2004ce}
A.~B. Zamolodchikov, ``{Expectation value of composite field T anti-T in
  two-dimensional quantum field theory},''
  \href{http://arxiv.org/abs/hep-th/0401146}{{\tt arXiv:hep-th/0401146}}.

\bibitem{Smirnov:2016lqw}
F.~A. Smirnov and A.~B. Zamolodchikov, ``{On space of integrable quantum field
  theories},'' \href{http://dx.doi.org/10.1016/j.nuclphysb.2016.12.014}{{\em
  Nucl. Phys. B} {\bf 915} (2017)  363--383},
  \href{http://arxiv.org/abs/1608.05499}{{\tt arXiv:1608.05499 [hep-th]}}.

\bibitem{giveon2020tbartlst}
A.~Giveon, N.~Itzhaki, and D.~Kutasov, ``{$ \mathrm{T}\overline{\mathrm{T}} $
  and LST},'' \href{http://dx.doi.org/10.1007/JHEP07(2017)122}{{\em JHEP} {\bf
  07} (2017)  122}, \href{http://arxiv.org/abs/1701.05576}{{\tt
  arXiv:1701.05576 [hep-th]}}.

\bibitem{Giveon:2017myj}
A.~Giveon, N.~Itzhaki, and D.~Kutasov, ``{A solvable irrelevant deformation of
  AdS$_{3}$/CFT$_{2}$},'' \href{http://dx.doi.org/10.1007/JHEP12(2017)155}{{\em
  JHEP} {\bf 12} (2017)  155}, \href{http://arxiv.org/abs/1707.05800}{{\tt
  arXiv:1707.05800 [hep-th]}}.

\bibitem{Lunin_2005}
O.~Lunin and J.~M. Maldacena, ``{Deforming field theories with U(1) x U(1)
  global symmetry and their gravity duals},''
  \href{http://dx.doi.org/10.1088/1126-6708/2005/05/033}{{\em JHEP} {\bf 05}
  (2005)  033}, \href{http://arxiv.org/abs/hep-th/0502086}{{\tt
  arXiv:hep-th/0502086}}.

\bibitem{Frolov_2005}
S.~Frolov, ``{Lax pair for strings in Lunin-Maldacena background},''
  \href{http://dx.doi.org/10.1088/1126-6708/2005/05/069}{{\em JHEP} {\bf 05}
  (2005)  069}, \href{http://arxiv.org/abs/hep-th/0503201}{{\tt
  arXiv:hep-th/0503201}}.

\bibitem{Araujo_2019}
T.~Araujo, E.~{\'O}. Colg{\'a}in, Y.~Sakatani, M.~M. Sheikh-Jabbari, and
  H.~Yavartanoo, ``{Holographic integration of $T \bar{T}$ {\textbackslash}{\&}
  $J \bar{T}$ via $O(d,d)$},''
  \href{http://dx.doi.org/10.1007/JHEP03(2019)168}{{\em JHEP} {\bf 03} (2019)
  168}, \href{http://arxiv.org/abs/1811.03050}{{\tt arXiv:1811.03050
  [hep-th]}}.

\bibitem{Sadri:2003bk}
D.~Sadri and M.~M. Sheikh-Jabbari, ``{String theory on parallelizable pp
  waves},'' \href{http://dx.doi.org/10.1088/1126-6708/2003/06/005}{{\em JHEP}
  {\bf 06} (2003)  005}, \href{http://arxiv.org/abs/hep-th/0304169}{{\tt
  arXiv:hep-th/0304169}}.

\bibitem{Asrat:2023yzy}
M.~Asrat, ``{Moving holographic boundaries},''
  \href{http://dx.doi.org/10.1016/j.nuclphysb.2024.116699}{{\em Nucl. Phys. B}
  {\bf 1008} (2024)  116699}, \href{http://arxiv.org/abs/2305.15744}{{\tt
  arXiv:2305.15744 [hep-th]}}.

\bibitem{BUSCHER198759}
T.~H. Buscher, ``{A Symmetry of the String Background Field Equations},''
  \href{http://dx.doi.org/10.1016/0370-2693(87)90769-6}{{\em Phys. Lett. B}
  {\bf 194} (1987)  59--62}.

\bibitem{Itsios_2012}
G.~Itsios, Y.~Lozano, E.~O~Colgain, and K.~Sfetsos, ``{Non-Abelian T-duality
  and consistent truncations in type-II supergravity},''
  \href{http://dx.doi.org/10.1007/JHEP08(2012)132}{{\em JHEP} {\bf 08} (2012)
  132}, \href{http://arxiv.org/abs/1205.2274}{{\tt arXiv:1205.2274 [hep-th]}}.

\bibitem{Meessen_1999}
P.~Meessen and T.~Ortin, ``{An Sl(2,Z) multiplet of nine-dimensional type II
  supergravity theories},''
  \href{http://dx.doi.org/10.1016/S0550-3213(98)00780-9}{{\em Nucl. Phys. B}
  {\bf 541} (1999)  195--245}, \href{http://arxiv.org/abs/hep-th/9806120}{{\tt
  arXiv:hep-th/9806120}}.

\bibitem{Hashimoto:1999ut}
A.~Hashimoto and N.~Itzhaki, ``{Noncommutative Yang-Mills and the AdS / CFT
  correspondence},''
  \href{http://dx.doi.org/10.1016/S0370-2693(99)01037-0}{{\em Phys. Lett. B}
  {\bf 465} (1999)  142--147}, \href{http://arxiv.org/abs/hep-th/9907166}{{\tt
  arXiv:hep-th/9907166}}.

\bibitem{Maldacena:1999mh}
J.~M. Maldacena and J.~G. Russo, ``{Large N limit of noncommutative gauge
  theories},'' \href{http://dx.doi.org/10.1088/1126-6708/1999/09/025}{{\em
  JHEP} {\bf 09} (1999)  025}, \href{http://arxiv.org/abs/hep-th/9908134}{{\tt
  arXiv:hep-th/9908134}}.

\bibitem{Nastase2023}
H.~Nastase and J.~Sonnenschein, ``{TT{\textasciimacron} deformations and the
  pp-wave correspondence},''
  \href{http://dx.doi.org/10.1103/PhysRevD.108.026012}{{\em Phys. Rev. D} {\bf
  108} (2023) no.~2, 026012}, \href{http://arxiv.org/abs/2207.02257}{{\tt
  arXiv:2207.02257 [hep-th]}}.

\bibitem{Pasti_1997}
P.~Pasti, D.~P. Sorokin, and M.~Tonin, ``{On Lorentz invariant actions for
  chiral p forms},'' \href{http://dx.doi.org/10.1103/PhysRevD.55.6292}{{\em
  Phys. Rev. D} {\bf 55} (1997)  6292--6298},
  \href{http://arxiv.org/abs/hep-th/9611100}{{\tt arXiv:hep-th/9611100}}.

\bibitem{Page:1983mke}
D.~N. Page, ``{Classical Stability of Round and Squashed Seven Spheres in
  Eleven-dimensional Supergravity},''
  \href{http://dx.doi.org/10.1103/PhysRevD.28.2976}{{\em Phys. Rev. D} {\bf 28}
  (1983)  2976}.

\bibitem{marolf2000chernsimonstermsnotionscharge}
D.~Marolf, ``{Chern-Simons terms and the three notions of charge},'' in {\em
  {International Conference on Quantization, Gauge Theory, and Strings:
  Conference Dedicated to the Memory of Professor Efim Fradkin}}, pp.~312--320.
\newblock 6, 2000.
\newblock \href{http://arxiv.org/abs/hep-th/0006117}{{\tt
  arXiv:hep-th/0006117}}.

\bibitem{ferko2025holographynullboundaries}
C.~Ferko and S.~Sethi, ``{Holography with Null Boundaries},''
  \href{http://arxiv.org/abs/2506.20765}{{\tt arXiv:2506.20765 [hep-th]}}.

\bibitem{Delduc_2014}
F.~Delduc, M.~Magro, and B.~Vicedo, ``{An integrable deformation of the $AdS_5
  \times S^5$ superstring action},''
  \href{http://dx.doi.org/10.1103/PhysRevLett.112.051601}{{\em Phys. Rev.
  Lett.} {\bf 112} (2014) no.~5, 051601},
  \href{http://arxiv.org/abs/1309.5850}{{\tt arXiv:1309.5850 [hep-th]}}.

\bibitem{Aharony:1998ub}
O.~Aharony, M.~Berkooz, D.~Kutasov, and N.~Seiberg, ``{Linear dilatons, NS
  five-branes and holography},''
  \href{http://dx.doi.org/10.1088/1126-6708/1998/10/004}{{\em JHEP} {\bf 10}
  (1998)  004}, \href{http://arxiv.org/abs/hep-th/9808149}{{\tt
  arXiv:hep-th/9808149}}.

\bibitem{Penrose1976}
R.~Penrose, ``Any spacetime has a plane wave as a limit,'' in {\em Differential
  Geometry and Relativity}, M.~Cahen and M.~Flato, eds., pp.~271--275.
\newblock Reidel, Dordrecht, 1976.

\bibitem{do2016differential}
M.~do~Carmo, {\em Differential Geometry of Curves and Surfaces: Revised and
  Updated Second Edition}.
\newblock Dover Books on Mathematics. Dover Publications, 2016.
\newblock \url{https://books.google.com.br/books?id=v4vqjgEACAAJ}.

\bibitem{Polchinski_1995}
J.~Polchinski, ``{Dirichlet Branes and Ramond-Ramond charges},''
  \href{http://dx.doi.org/10.1103/PhysRevLett.75.4724}{{\em Phys. Rev. Lett.}
  {\bf 75} (1995)  4724--4727}, \href{http://arxiv.org/abs/hep-th/9510017}{{\tt
  arXiv:hep-th/9510017}}.

\bibitem{POLYAKOV1981207}
A.~M. Polyakov, ``{Quantum Geometry of Bosonic Strings},''
  \href{http://dx.doi.org/10.1016/0370-2693(81)90743-7}{{\em Phys. Lett. B}
  {\bf 103} (1981)  207--210}.

\bibitem{Duval1991}
C.~Duval, G.~W. Gibbons, and P.~Horvathy, ``{Celestial mechanics, conformal
  structures and gravitational waves},''
  \href{http://dx.doi.org/10.1103/PhysRevD.43.3907}{{\em Phys. Rev. D} {\bf 43}
  (1991)  3907--3922}, \href{http://arxiv.org/abs/hep-th/0512188}{{\tt
  arXiv:hep-th/0512188}}.

\bibitem{Seiberg_1999}
N.~Seiberg and E.~Witten, ``{String theory and noncommutative geometry},''
  \href{http://dx.doi.org/10.1088/1126-6708/1999/09/032}{{\em JHEP} {\bf 09}
  (1999)  032}, \href{http://arxiv.org/abs/hep-th/9908142}{{\tt
  arXiv:hep-th/9908142}}.

\bibitem{Imeroni_2008}
E.~Imeroni, ``{On deformed gauge theories and their string/M-theory duals},''
  \href{http://dx.doi.org/10.1088/1126-6708/2008/10/026}{{\em JHEP} {\bf 10}
  (2008)  026}, \href{http://arxiv.org/abs/0808.1271}{{\tt arXiv:0808.1271
  [hep-th]}}.

\bibitem{Araujo:2017enj}
T.~Araujo, E.~{\'O}~Colg{\'a}in, J.~Sakamoto, M.~M. Sheikh-Jabbari, and
  K.~Yoshida, ``{$I$ in generalized supergravity},''
  \href{http://dx.doi.org/10.1140/epjc/s10052-017-5316-5}{{\em Eur. Phys. J. C}
  {\bf 77} (2017) no.~11, 739}, \href{http://arxiv.org/abs/1708.03163}{{\tt
  arXiv:1708.03163 [hep-th]}}.

\bibitem{Filev:2007gb}
V.~G. Filev, C.~V. Johnson, R.~C. Rashkov, and K.~S. Viswanathan, ``{Flavoured
  large N gauge theory in an external magnetic field},''
  \href{http://dx.doi.org/10.1088/1126-6708/2007/10/019}{{\em JHEP} {\bf 10}
  (2007)  019}, \href{http://arxiv.org/abs/hep-th/0701001}{{\tt
  arXiv:hep-th/0701001}}.

\end{thebibliography}\endgroup

\end{document}